\documentclass[]{agujournal}
\draftfalse
\usepackage{rotating}
\usepackage{longtable}

\usepackage{ulem}
\usepackage{hyperref}

\usepackage{xcolor}
\definecolor{hred}{RGB}{235,0,0}

\justifying

\journalname{Science Advances}

\newcounter{movie}

\begin{document}

\title{Giant impacts stochastically change the internal pressures of terrestrial planets}

\authors{Simon J. Lock\affil{1,2}, Sarah T. Stewart\affil{3}}

\affiliation{1}{Division of Geological and Planetary Sciences, Caltech}
\affiliation{2}{Department of Earth and Planetary Sciences, Harvard University}
\affiliation{3}{Department of Earth and Planetary Sciences, U. California Davis}

\correspondingauthor{Simon J. Lock}{slock@caltech.edu}

\begin{keypoints}

\item The pressures inside terrestrial planets are stochastically perturbed due to giant impacts, which requires changes in the interpretation of geochemical tracers of accretion.

\end{keypoints}

\begin{abstract} 

Pressure is a key parameter in the physics and chemistry of planet formation and evolution. Previous studies have erroneously assumed that internal pressures monotonically increase with the mass of a body. Using smoothed particle hydrodynamics and potential field method calculations, we demonstrate that the hot, rapidly-rotating bodies produced by giant impacts can have much lower internal pressures than cool, slowly-rotating planets of the same mass. Pressures subsequently increase due to thermal and rotational evolution of the body. Using the Moon-forming impact as an example, we show that the internal pressures after the collision could have been less than half that in present-day Earth. The current pressure profile was not established until Earth cooled and the Moon receded, a process that may take up to 10s Myr after the last giant impact. Our work defines a new paradigm for pressure evolution during accretion of terrestrial planets: stochastic changes driven by impacts.


\end{abstract}

\section{Introduction}
\label{sec:intro}

\begin{figure}
\centering
\includegraphics[height=0.96\textheight]{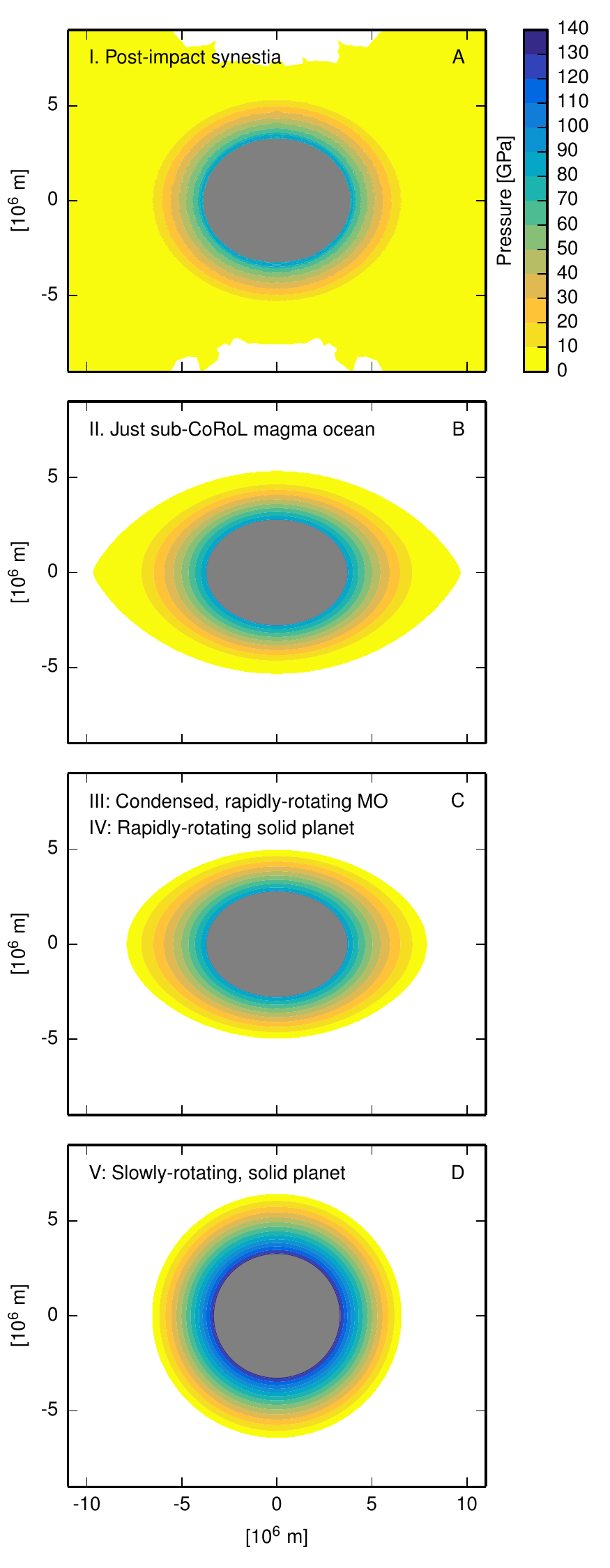}
\caption{Caption on next page.}
\end{figure}
\setcounter{figure}{0}
\begin{figure}
\caption{Cooling and tidal evolution after the Moon-forming giant impact forced the shape and internal pressures in Earth to change. This figure presents contours of the internal pressure in Earth at different stages in its evolution after an impact of two 0.52~$M_{\rm Earth}$ bodies with an impact velocity of 9.7~km~s$^{-1}$ and an impact parameter of 0.55. The post-impact body has a mass of 0.97~$M_{\rm Earth}$ and an angular momentum of 2.16~$L_{\rm EM}$. In this example, Earth was initially ($\sim$~days after impact) a synestia with modest internal pressures and a large moment of inertia (A). Subsequent cooling led to condensation of the silicate vapor and the body first fell below the corotation limit (B) then cooled to magma-ocean (MO) planet with a volatile-dominated atmosphere (C). The planet was rotating rapidly and the internal pressures remained modest. Continued cooling solidified the magma ocean but had little effect on the shape of the planet or on the internal pressures. Tidal recession of the Moon to the point that the lunar spin axis underwent the Cassini state transition reduced the angular momentum of Earth, the planet became spherical, and the internal pressure significantly increased (D).
}
\label{fig:cartoon} 
\end{figure}

The end of the main stage of terrestrial planet formation is characterized by high-energy collisions between planetary bodies, called giant impacts. Giant impacts melt and vaporize the silicate mantles of the impacting bodies \cite{Nakajima2015,Lock2017}, and planets can acquire significant angular momentum (AM) via one or more giant impacts \cite{Kokubo2010}. Through multiple giant and smaller impacts, the terrestrial planets in our solar system grew from roughly Mars-mass planetary embryos to nearly their final masses over a few to a few tens of millions of years.  

Internal pressures in proto-planets change during formation. Internal pressures control a number of key processes that produce geochemical signatures used to understand the mechanisms and timescales of accretion. Metal-silicate equilibration and segregation of iron to the core likely occurs after giant impacts when the lower mantles of post-impact bodies are substantially molten \cite{Rubie2007}. Partitioning of elements between the mantle and core is strongly dependent on the pressure and temperature of equilibration \cite[e.g.,][]{Li1996,Righter1997}. The pressure and temperature profiles of post-impact bodies hence determine the distribution of elements between the core and the mantle. The partitioning of elements affects chemical and isotopic systems that are used as tracers for planetary accretion, such as the moderately siderophile elements (MSEs) \cite[e.g.,][]{Li1996,Righter1997,Yu2011a,Rubie2015a,Piet2017,Fischer2017} and the Hf-W system \cite[e.g.,][]{Jacobsen2005}. To use these geochemical systems to trace the process of planet formation, we need to understand the evolution of internal pressures and temperatures during accretion.

How a magma ocean freezes, whether from the bottom up or middle out, depends on the pressure profile in the post-impact body because the relative slopes of adiabats and the phase boundary change with pressure. At modest pressures, the temperature of the liquid adiabat increases less rapidly with pressure than the liquidus for a mantle of bulk silicate Earth (BSE) composition, and smaller bodies begin to freeze from the core-mantle boundary (CMB) outwards. However, the slopes of liquid adiabats and the liquidus likely cross at high pressure, and it has been calculated that the liquid mantle adiabat first intersects the liquidus at pressures of around 70 to 105~GPa \cite{Stixrude2009,Thomas2013}. Due to iron enrichment in the liquid, melt near the intersection could be of a similar density to the solid and might even be denser at higher pressures \cite{Thomas2013}. In bodies with a CMB pressure higher than the intersection, the mantle would begin to freeze from the mid-mantle, potentially isolating a basal magma ocean in the lower mantle \cite{Labrosse2007}. If the internal pressures during freezing of its terminal magma ocean were similar to the present day, Earth may have had a basal magma ocean. The chemical heterogeneity produced by crystallization of a magma ocean, potentially even persistence of ancient melt, has been proposed \cite[e.g.,][]{Labrosse2007,Williams1996,Wicks2017} as a mechanism to explain the dense, seismically anomalous regions observed in the lower mantle today \cite[e.g.,][]{Trampert2004,Ishii1999}. Also, a basal magma ocean would have had consequences for Earth's magnetic dynamo \cite[e.g.,][]{Ziegler2013}.

Current models of accretion assume that internal pressures are dependent only on mass and monotonically increase as planets grow \cite[e.g.,][]{Yu2011a,Rubie2015a,Fischer2017}. Internal pressures have been calculated using simple interior models of condensed (liquid or solid), non-rotating bodies. However, giant impacts can radically alter the shape, thermal state and AM of terrestrial bodies \cite{Lock2017,Lock2018moon}. The shock energy deposited by a giant impact is great enough to transform the outer 10s of percent of the body to vapor or super-critical fluid. In addition, the centrifugal force in rotating bodies acts against gravity and causes flattening perpendicular to the rotation axis. In many giant impacts, the post-impact body exceeds the corotation limit (CoRoL) due to a combination of these two effects \cite{Lock2017}. The CoRoL is defined by where the angular velocity at the equator of a corotating planet intersects the Keplerian orbital velocity. The CoRoL is a surface that depends on thermal state, AM, total mass, and compositional layering. Fluid bodies that exceed the CoRoL are named {\it synestias}. Post-impact bodies evolve rapidly due to cooling and condensation of the silicate vapor \cite{Lock2018moon,Lock2018LPSC} and changes in AM driven by tidal interactions between satellites and the Sun. The dramatic changes in physical structure (e.g., shape, mass and AM distribution, pressure and temperature profiles) induced by impacts, and evolution in structure during the subsequent recovery, have not been examined. Changes in physical structure have implications for the interpretation of geochemical tracers of accretion, and study of these processes is essential for understanding terrestrial planet formation.

Here, we investigate the effects that changes in shape, thermal state and AM have on internal pressures, with a focus on potential Moon-forming giant impacts. The advantage of studying potential Moon-forming impacts is that there are strong observational constraints from the Earth-Moon system. The canonical Moon-forming impact is a relatively low energy impact that left the Earth-Moon system with close to its present AM, $L_{\rm EM} = 3.5 \times 10^{34}$~kg~m$^{2}$~s$^{-1}$ \cite{Canup2004}. But, in recent years, several mechanisms have been found that could have transferred AM away from the Earth-Moon system through three-body interactions between the Sun, Earth, and Moon \cite{Cuk2012,Wisdom2015,Cuk2016}. This discovery has increased the range of possible Moon-forming impacts, and a range of high-energy, high-AM collisions have been proposed as potential candidates \cite{Cuk2012,Canup2012,Lock2018moon}. The specific energies of such impacts are typically an order of magnitude greater than the canonical impact and the AM ranges from 1.8 to 3.2~$L_{\rm EM}$. The last giant impact sets the conditions for the subsequent evolution of Earth and there is a need to find observational tests to differentiate between these scenarios for lunar origin. Here, we calculate the effect of different Moon-forming impacts on internal pressures and examine the implications for the physical and chemical properties of Earth.

It is not yet feasible to construct a dynamical model of how Earth transitioned from a hot, rapidly-rotating post-impact body to the solid, slowly rotating planet we know today. Here, we take the approach of comparing the internal pressures at five different stages in Earth's recovery after the Moon-forming giant impact: (I) immediately after the impact; (II) once the body cooled to below the CoRoL; (III) once the silicate vapor fully condensed to a magma ocean; (IV) after magma ocean solidification; and (V) after tidal recession of the Moon to the Cassini-state transition. Figure~\ref{fig:cartoon} shows internal pressure contours at each of these stages after a high-energy, high-AM Moon-forming impact. We present calculations of the internal pressures in Earth at each of these stages and discuss the mechanisms driving changes in pressure between stages.

\section{Results}
\label{sec:results}

\subsection{Stage I: Immediately after a giant impact}
\label{sec:results:PI}

\begin{figure*}
\centering
\includegraphics{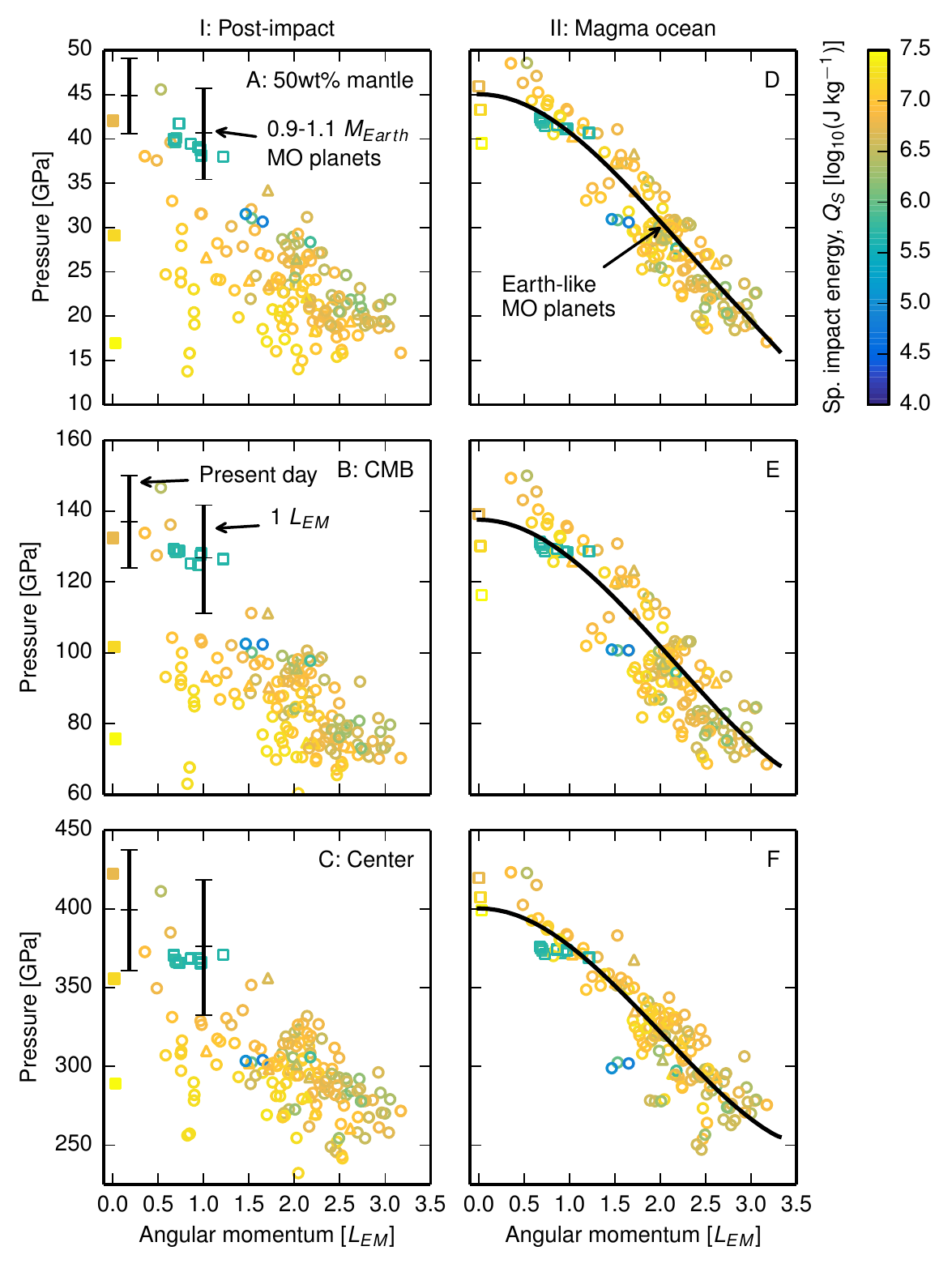}
\caption{Pressures in the interior of Earth-mass bodies can be substantially lower after giant impacts than in the modern Earth. The pressures in the middle of the mantle by mass (A), at the core-mantle boundary (B), and at the center (C) of post-impact bodies are presented as a function of the angular momentum of the bound mass (symbols). Symbols indicate structures that are above ($\circ$), below ($\Box$), or have an unclear relationship to ($\bigtriangleup$) the corotation limit \cite{Lock2017}. The post-impact bodies plotted are restricted to those with a bound mass between 0.9 and 1.1 $M_{\rm Earth}$, and colors indicate the geometrically-modified specific energy of the impact, $Q_{\rm S}$. In A-C, black bars denote the range of pressures in magma-ocean (MO) planets of between 0.9 and 1.1 $M_{\rm Earth}$ and angular momenta equal to that of either the present-day Earth-Moon system (1~$L_{EM}$) or the present-day Earth alone (about 0.18~$L_{EM}$). The pressures at the same levels are shown for magma-ocean planets of the same mass, angular momenta and core-mass fraction as each of the post-impact bodies (D-F, symbols). The black line describes the pressure in an Earth-like magma-ocean planet as a function of angular momentum calculated using the HERCULES planetary-structure code. Filled symbols are discussed in text.}
\label{fig:press_evo_abs}
\end{figure*}

\begin{figure}
\centering
\includegraphics{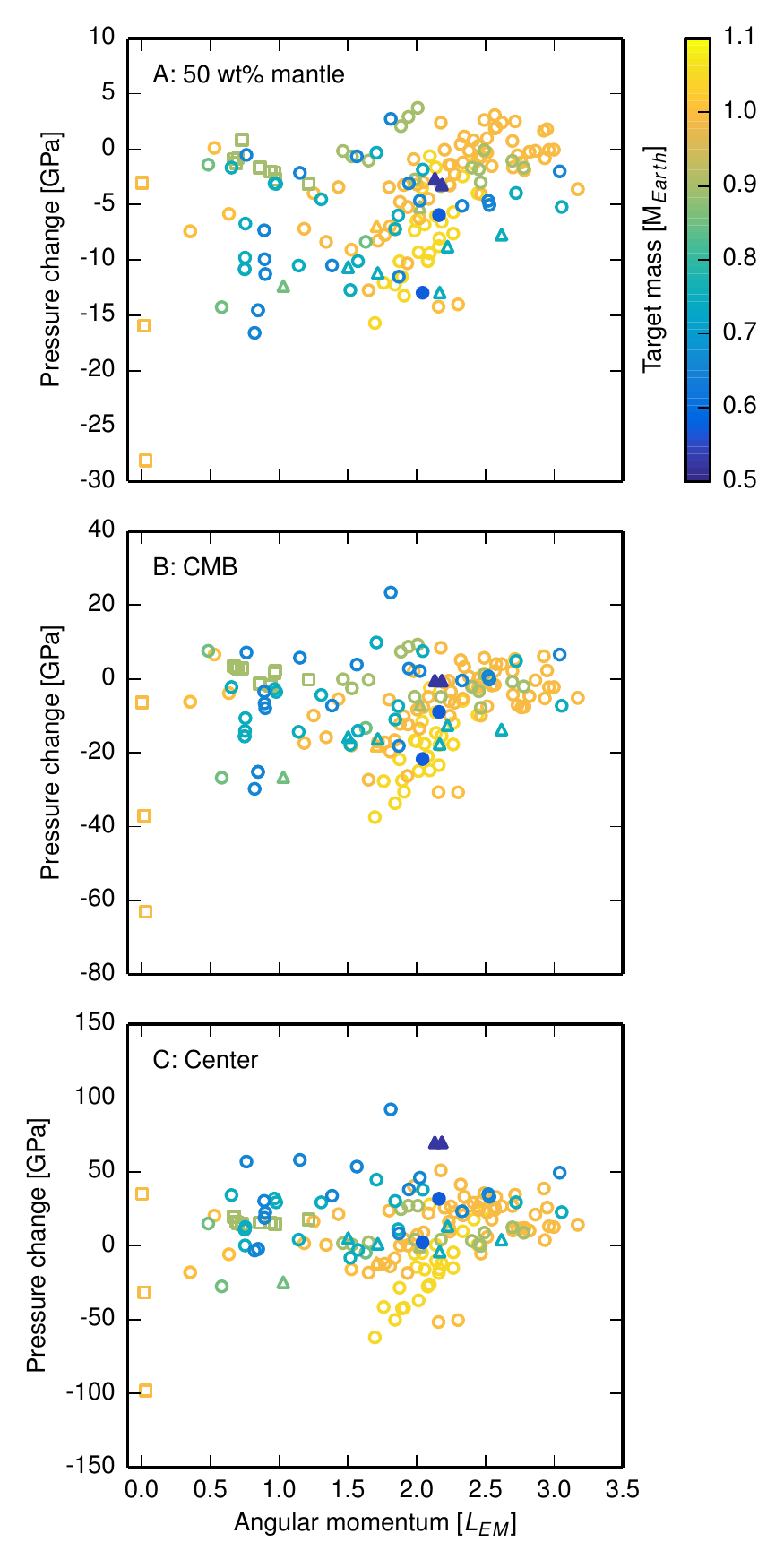}
\caption{Contrary to expectations, the pressures in bodies after giant impacts are often lower than that in the larger of the pre-impact bodies. Symbols show the pressure differences between the pre-impact target and post-impact body for the impacts in Figure~\ref{fig:press_evo_abs}A-C. The pressure in the target bodies were calculated using HERCULES, assuming the body was a magma-ocean planet. Panels show the pressure difference at the middle of the mantle by mass (A), the core-mantle boundary (B), and the center (C) of bodies as a function of the angular momentum of the post-impact bound mass. Colors indicate the mass of the target body and symbols are the same as in Figure~\ref{fig:press_evo_abs}. Filled symbols are discussed in text.}
\label{fig:target_press} 
\end{figure}

We find that the pressures within Earth-mass bodies can be much lower after giant impacts than in the present-day Earth. Figure~\ref{fig:press_evo_abs} shows the pressure in the middle of the mantle by mass (A), at the CMB (B), and at the center (C) of post-impact bodies produced by smoothed particle hydrodynamics (SPH) giant impact simulations (Section~\ref{sup:sec:methods:SPH}) when the body has reached gravitational equilibrium ($\sim48$~hrs after the impact, Figure~\ref{fig:cartoon}A). We simulated collisions with a large range of impact parameters that created approximately Earth-mass final bodies, and the bodies shown are not restricted to those formed by potential Moon-forming impacts. Only bodies with final bound masses between 0.9 and $1.1$ Earth masses, $M_{\rm Earth}$, are shown as a function of the AM of the post-impact bound mass. Colors indicate the modified specific energy, $Q_{\rm S}$ (Equation~\ref{sup:eqn:QS}), of the impacts that generated each structure. The specific entropy of the outer layers of post-impact bodies scales well with $Q_{\rm S}$ \cite{Lock2017}. For reference, the black bars show the range of pressures for magma-ocean planets with masses of 0.9, 1 and $1.1 M_{\rm Earth}$, and AM corresponding to that of the present-day Earth ($\sim0.18$~$L_{\rm EM}$) and the total AM of the present-day Earth-Moon system.

The lower pressures in the interior of post-impact bodies are due to a combination of factors. First, for rotating bodies the pressure is decreased due to the fictitious centrifugal force. The centrifugal force acts against gravity in the direction perpendicular to the rotation axis, leading to a shallower pressure gradient. The faster a body rotates, the lower the pressures. A general trend of decreasing pressure with increasing AM can be seen in Figure~\ref{fig:press_evo_abs}A-C. The effect of rotation on internal pressures can be seen more clearly for bodies with the same thermal state. For example, Figure~\ref{fig:press_evo_abs}D-F shows the internal pressures in traditional magma-ocean planets, bodies with molten mantles and volatile-dominated atmospheres. Here, we define our fiducial magma-ocean planets as bodies with isentropic mantles with a 10~bar potential temperature of 4000~K and isentropic cores with a temperature of $\sim3800$ K at the pressure of the present-day CMB, similar to the present-day Earth. The points show the pressures in magma-ocean planets, calculated using the HERCULES planetary-structure code \cite{Lock2017}, of the same mass, composition and AM as the various post-impact bodies shown in Figure~\ref{fig:press_evo_abs}A-C. The black lines show the pressures in an Earth-mass, Earth-composition planet as a function of AM.  For a body with a constant thermal state, increases in angular momentum are accompanied by dramatic decreases in internal pressures. We discuss the specific case of magma-ocean planets in more detail in Section~\ref{sec:results:cond}.

Second, the bulk densities of post-impact bodies are much lower than fully condensed planets due to their hot thermal state and high vapor fraction. Decreasing the density of a body lowers the pressures within the body. The effect of bulk density can be demonstrated by a simple calculation. Consider a constant-density, non-rotating body of a given mass. The pressure is given by the integral of the pressure gradient from the surface. Assuming hydrostatic equilibrium,
\begin{equation}
p(r)=\int_a^r \frac{\mathrm{d}p}{\mathrm{d}r'} \mathrm{d}r' = - \int_a^r \rho g(r') \mathrm{d}r' \; ,
\label{eqn:hydrostatic}
\end{equation}
where $p$ is pressure, $a$ is the radius of the body, $r$ is the radius at a point in the body, $\rho$ is the density, and $g(r)$ is the gravitational acceleration at radius $r$. We assume zero pressure at the surface. The gravitational acceleration in a spherical, constant density body is given by
\begin{equation}
g(r) = \frac{G M_{r'<r}}{r^2} = \frac{4 \pi G \rho r }{3} \; ,
\label{eqn:gravity}
\end{equation}
where $G$ is the gravitational constant, and $M_{r'<r}$ is the mass interior to the radius $r$. Integrating Equation~\ref{eqn:hydrostatic} with Equation~\ref{eqn:gravity} gives internal pressures of
\begin{equation}
p(r) = \frac{2 \pi \rho^2 G}{3} \left ( a^2 - r^2 \right ) \; .
\label{eqn:p_r}
\end{equation}
The radius which encloses a given mass fraction, $f$, is
\begin{equation}
r(f) = \left [ \frac{3 f M }{4 \pi \rho} \right ]^{\frac{1}{3}} \; ,
\end{equation}
where $M$ is the total mass of the body. Rewriting Equation~\ref{eqn:p_r} as a function of $f$ gives
\begin{equation}
p(f) = \frac{1}{2} \left ( \frac{4 \pi}{3} \right )^{\frac{1}{3}} G M^{\frac{2}{3}} \rho^{\frac{4}{3}} \left [1- f^{\frac{2}{3}}  \right ] \; .
\end{equation}
If the density of a body decreases then the pressure of any given mass fraction also decreases. This effect is well demonstrated in Figure~\ref{fig:press_evo_abs}A-C by a series of three head-on impacts of increasing impact energy (filled yellow squares with near-zero AM). Two impacts of 0.05~$M_{\rm Earth}$ bodies onto 0.99~$M_{\rm Earth}$ at velocities of 15 and 25~km~s$^{-1}$ produce bodies with CMB pressures of 132 and 101~GPa, respectively. A higher energy impact of a 0.1~$M_{\rm Earth}$ body at 25~km~s$^{-1}$ produces a body with a CMB pressure of 76~GPa. The corresponding modified specific energies are $Q_{\rm S}=5$, 15 and 28~MJ~kg$^{-1}$.  As the impact energy increases, post-impact bodies are more vaporized and hence have lower bulk density. The result is a substantial decrease in pressure with increasing impact energy. 

The trends with both impact energy and AM in Figure~\ref{fig:press_evo_abs}A-C are complicated by the fact that the bulk density and rotation rate are not independent. Generally, the lower the density of a body, the greater its spatial extent and the higher its moment of inertia. A lower density body of the same AM will rotate more slowly and so reduce the strength of the centrifugal force. The balance between rotation and bulk density controls the internal pressures. The pressures depend on the precise post-impact structure and hence the parameters of the impact.

We find that the pressures in the mantles of post-impact bodies are often lower than in the larger of the pre-impact bodies (Figure ~\ref{fig:target_press}). For example, there are four post-impact bodies that had target masses of 0.52 or 0.57~$M_{\rm Earth}$ (filled symbols in Figure~\ref{fig:target_press}) that, despite almost doubling in mass, have post-impact CMB pressures that are the same as or significantly less (by up to $\sim$20~GPa) than in the target. This result is opposite to the previous paradigm of internal pressures increasing with increasing mass through accretion.

The post-impact pressures vary substantially between different Moon-formation models. Sub-CoRoL, post-impact bodies produced by low-energy, low-AM collisions similar to the canonical impact (green squares with AM $\sim 1 L_{\rm EM}$ in Figure~\ref{fig:press_evo_abs}) have modest rotation rates (periods $\gtrsim 5$~hr) and vapor fractions. The effect of lower density and rotation are not substantial and the internal pressures are comparable to magma-ocean planets of the same AM (black bars in Figure~2) and only somewhat lower than those in present-day Earth. However, the synestias produced by high-energy, high-AM impacts, such as those invoked in more recent lunar origin models \cite{Cuk2012,Canup2012,Lock2018moon}, can have much lower bulk densities and rapid rotation rates. The combined effect is much lower internal pressures throughout the bodies. For example, the CMB pressure can be as low as about 60~GPa, less than half of the present Earth value of $\sim 136$~GPa.

\subsection{Stages II and III: Condensation of the silicate vapor}
\label{sec:results:cond}

\begin{sidewaysfigure*}
\centering
\includegraphics{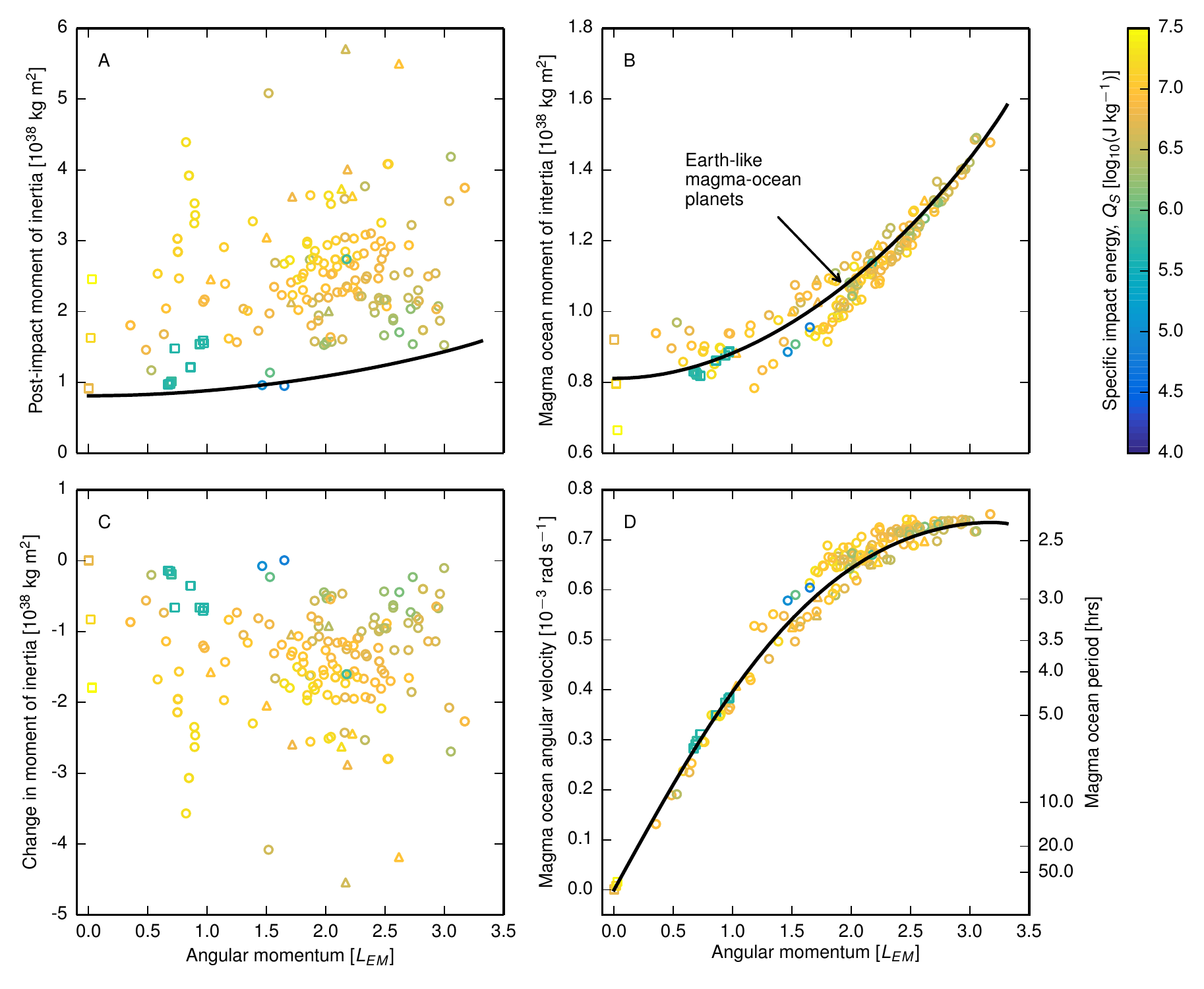}
\caption{Caption on next page.}
\end{sidewaysfigure*}
\setcounter{figure}{3}
\begin{figure*}
\caption{The moment of inertia of condensed magma-ocean planets is lower, and their rotation rates faster, than those of post-impact bodies. (A) The moment of inertia of the post-impact bodies shown in Figure~\ref{fig:press_evo_abs}. (B) The moment of inertia of magma-ocean planets with the same mass, composition and angular momenta. (C) The change in the moment of inertia due to condensation of the silicate vapor. (D) The angular velocity of magma-ocean planets with the same mass, composition and angular momenta. Colors and symbols are the same as in Figure~\ref{fig:press_evo_abs}. Black lines show the moment of inertia (A, B) or angular velocity (D) of an Earth-like, magma-ocean planet as a function of angular momentum. Note the large change in vertical scale between A and B.
}
\label{fig:omgchange_cooling} 
\end{figure*}

\begin{figure*}
\centering
\includegraphics{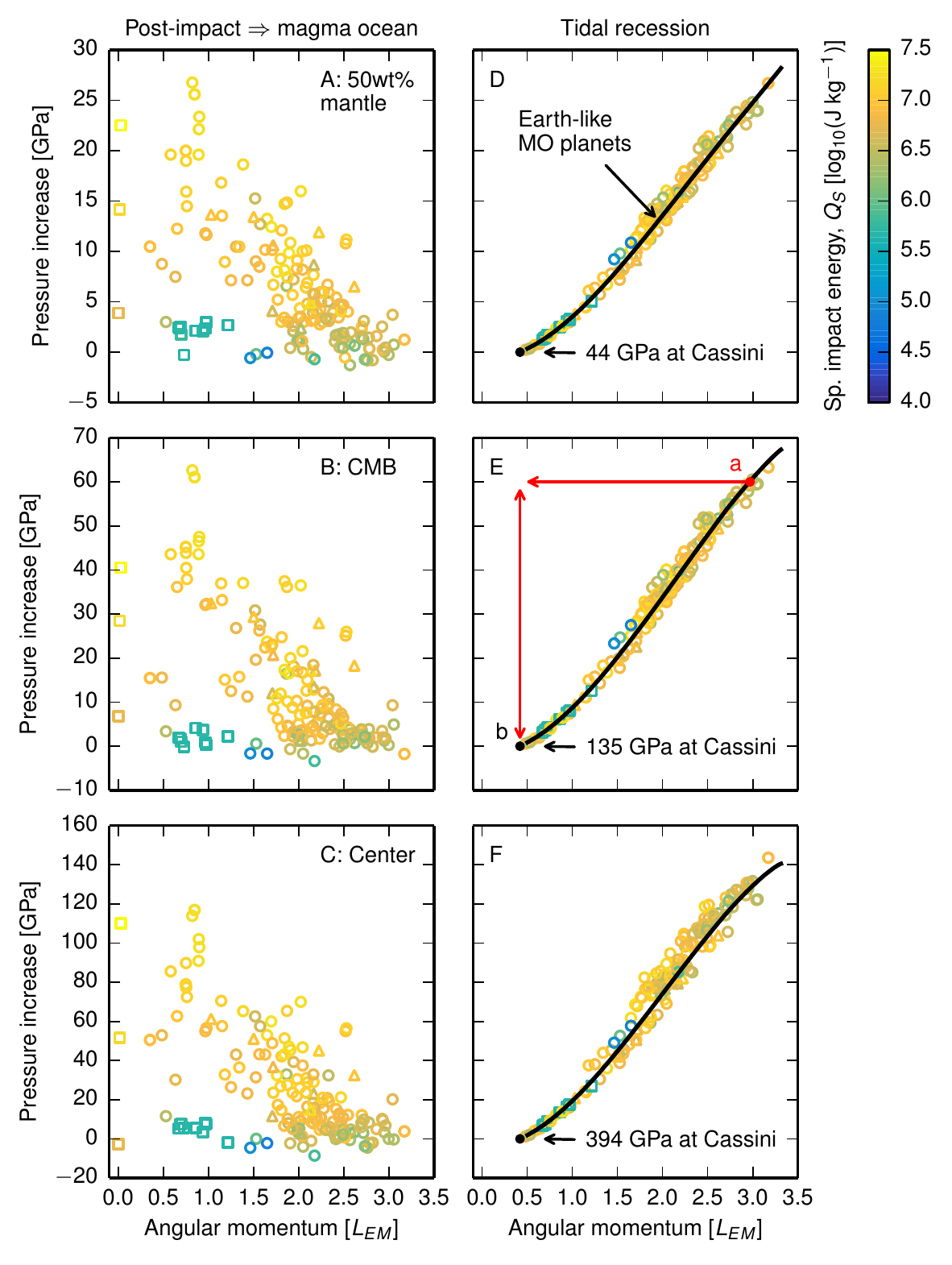}
\caption{Pressures in the interior of Earth-mass bodies change substantially during cooling and tidal evolution. The change upon condensation of the post-impact vapor, i.e., the difference in pressure between the post-impact state (Figure~\ref{fig:press_evo_abs}A-C) and the magma-ocean (MO) planet (Figure~\ref{fig:press_evo_abs}D-F), varies substantially between different impacts (A-C). The decrease in the angular momentum of Earth during lunar tidal recession could have substantially increased the internal pressures in the planet. D-F demonstrate the increase in internal pressures upon tidal recession of the Moon to the Cassini state transition for a body with a given initial angular momentum. Assuming the angular momentum of the Earth-Moon system had reached its present-day value, the angular momentum of Earth at the Cassini-state transition (a lunar semi-major axis of about 30 Earth radii) was 0.417~$L_{\rm EM}$. The black line shows the pressure increase for an Earth-like body. Symbols and colors are the same as in Figure~\ref{fig:press_evo_abs}. Paths and points in E are discussed in text.}
\label{fig:press_evo_change}
\end{figure*}

The substantially vaporized and extended post-impact bodies rapidly evolve due to radiative cooling. After most impacts the mantle of a post-impact body transitions smoothly from vapor to supercritical fluid to liquid at high pressure, and there is no liquid surface overlain by a silicate atmosphere \cite{Lock2017,Stewart2018LPSC}. The photosphere is dominated by droplets of liquid silicate and the body is radiating at about 2300~K \cite{Lock2018moon}. Such high radiative temperatures drive rapid condensation of the silicate vapor over a few hundred to thousands of years \cite{Lock2018LPSC}. If the body is initially above the CoRoL, the synestia first cools to become a corotating planet (Figure~\ref{fig:cartoon}B, stage II). Continued cooling leads to the condensation of the remaining silicate vapor and the body becomes a traditional magma-ocean planet with a fully liquid upper mantle and a volatile-dominated atmosphere (Figure~\ref{fig:cartoon}C, stage III). The magma-ocean planet can be substantially oblate due to its rapid rotation. After canonical Moon-forming impacts, the equatorial radius of the planet is $\sim10$ \% larger than the polar radius, but after some high-AM impacts the equatorial radius is twice as large as the polar radius. 

Condensation of the silicate vapor substantially reduces the size of the post-impact body, increasing the bulk density and changing the interior pressures. Figure~\ref{fig:press_evo_abs}D-F shows the internal pressure in the magma ocean after complete condensation of the silicate vapor atmosphere. Due to the effects of rotation, the pressures in Earth-like magma-ocean planets are typically lower than in the present-day Earth. The rotation rate of magma-ocean planets produced by canonical Moon-forming giant impacts are modest (about 5~hr period), the centrifugal force is small, and the internal pressures are within $\sim$10~GPa of those in the present-day Earth. In high-AM scenarios, the rotation rates are much higher and the internal pressures are lower than the present day. For example, the CMB pressure in the range of AM proposed for high-AM impacts ($\geq 1.8 L_{\rm EM}$) is less than about 110~GPa and can be as low as 70~GPa. 

The change in pressure due to condensation is a balance between the competing effects of changes in rotation rate and an increase in bulk density. As a result, the pressure change is highly variable. The condensed planet is rotating more rapidly than most of the material in the post-impact body. The angular velocity, $\omega$, of a parcel of material around the center of mass of a body is given by
\begin{equation}
\omega = \frac{\delta L}{\delta I} \; ,
\label{eqn:omega_gen}
\end{equation}
where $\delta L $ is the AM of the parcel about the center of mass of the body, $\delta I = r_{xy}^2 \delta m$ is the moment of inertia of the parcel about the center of mass, $\delta m$ is the mass of the parcel, and $r_{xy}$ is the distance from the rotation axis. For a corotating body, where $\omega$ is the same for all mass, $\omega$ can be expressed as
\begin{equation}
\omega = \frac{L}{I} \; ,
\label{eqn:omega}
\end{equation}
where $L$ is the total AM of the body, $I$ is its moment of inertia,
\begin{equation}
I = \int_{\mathcal{V}} \mathrm{d}I' = \int_{\mathcal{V}} r_{xy}^2 \mathrm{d}m' \; ,
\label{eqn:I}
\end{equation}
and $\mathcal{V}$ is the volume over which there is mass. The angular velocity in post-impact bodies varies with distance from the rotation axis with the outer regions rotating more slowly than the inner corotating region. More mass is further from the rotation axis than in an equivalent condensed body of the same mass and AM, and the total moment of inertia of the body, as given by Equation~\ref{eqn:I}, is significantly (by up to a factor of 6) higher (Figure~\ref{fig:omgchange_cooling}A). The mass farther from the rotation axis has high specific-AM and accommodates a significant fraction of the total AM. As a result, the angular velocity of most material in a post-impact body is lower than in an equivalent corotating planet (Figure~\ref{fig:omgchange_cooling}D).

Condensation of the silicate vapor substantially reduces the moment of inertia of a body (by a factor of several, Figure~\ref{fig:omgchange_cooling}B,C) and the total AM must be accommodated entirely in the corotating magma ocean. Consequently, the rotation rate of the magma-ocean planet after condensation (Figure~\ref{fig:omgchange_cooling}D) is faster than that of the material in the structure immediately after the impact. This fact was not appreciated in previous studies \cite[e.g.,][]{Rufu2017}. Since the mass distribution, and hence moment of inertia, of bodies can be altered so substantially by changes in thermal state and AM, total AM must be used to quantify the rotation state of planetary bodies and not spin period.

The increase in rotation rate acts to decrease the pressure in the body, but in almost all cases the effect of the density increase dominates, and the pressures are greater in the magma-ocean planet than immediately after the impact. Figure~\ref{fig:press_evo_change}A-C shows the change in pressure due to condensation of the vapor, i.e., the difference between the pressures at stage I and III shown in the first and second column of Figure~\ref{fig:press_evo_abs}. The pressure increase after canonical Moon-forming giant impacts is typically small (a few gigapascals, green squares with AM $\sim 1 L_{\rm EM}$), but after high-energy impacts the pressures can increase dramatically, e.g., by several 10s~GPa at the CMB.

For most of the high-AM bodies that are initially above the CoRoL, a substantial fraction of the vapor must condense for the body to fall below the CoRoL. The pressures at the CoRoL (stage II) for high-AM bodies are hence similar to those in the magma ocean (stage III) (Figure~\ref{sup:fig:pCoRoL}). For bodies with lower AM ($\lesssim 1.5$~$L_{\rm EM}$), the outer layers of the body must be exceedingly hot (high entropy) for the body to exceed the CoRoL. Hence, for the few lower-AM bodies that are initially above the CoRoL, the CoRoL is reached earlier during cooling, and there can still be a substantial change in pressure between the CoRoL and magma ocean stages.

In Figure~\ref{fig:press_evo_abs}D-F and Figure~\ref{fig:press_evo_change}A-C, we did not include the effect of the formation of the Moon on the internal pressures and made comparison to magma-ocean planets that included all the bound mass and AM of the post-impact body. Over the short time it takes to condense the silicate vapor (on the order of hundred to thousands years, \cite{Lock2018LPSC}), any satellite produced would remain close to the central body \cite{Cuk2012,Wisdom2015,Cuk2016,Touma1994}. The formation of a close-in, lunar-mass satellite only leads to modest changes in internal pressure of a few percent, e.g., up to 7~GPa at the CMB of a magma-ocean planet for a Moon closer than ten Earth radii (Figure~\ref{sup:fig:pchange_cooling_Moon}) and so we ignore this effect for the rest of the paper. 

\subsection{Stage IV: The planet after solidification of the mantle}
\label{sec:results:freezing}

After condensation of the vapor, the planet continues to cool and solidifies over a timescale of between 10's~kyr \cite{Lebrun2013} and 10's~Myr \cite{Zahnle2015} (Figure~\ref{fig:cartoon}C, stage IV). The pressure in a body does not change significantly due to magma-ocean solidification as the volume change upon freezing for silicates is small, particularly at high pressure. We calculate that varying the thermal state of an Earth-like, condensed body only changes the internal pressure by a few percent, e.g., up to 4~GPa at the CMB (Figure~\ref{sup:fig:HERCULES_therm}). The pressures in solidified planets are approximately the same as in equivalent magma-ocean bodies. 

After high-AM impacts, the mantle freeze in a lower pressure environment than in equivalent non-rotating bodies. The pressures in the mantle of Earth after a high-AM Moon-forming impact would have been much lower than after a canonical giant impact. High-AM Moon-forming impacts thus create a different pressure environment for solidification of the terrestrial magma ocean than that of the present-day Earth.

\subsection{Stage V: The slowly rotating planet after tidal evolution}
\label{sec:results:despin}

As the Moon tidally receded from Earth, the AM of the planet decreased and its shape approached a sphere (Figure~\ref{fig:cartoon}D, stage V). We define stage V as when the lunar spin axis underwent the Cassini state transition (which occurred when the lunar semi-major axis was about 30 Earth radii), at which point Earth had an AM of 0.417~$L_{\rm EM}$. The reduction of the centrifugal force during tidal recession, and the associated change in shape, leads to an increase in the internal pressures in the planet.

Although the exact AM history is debated \cite{Cuk2012,Wisdom2015,Cuk2016,Touma1994}, by the point the lunar spin axis underwent the Cassini state transition, the AM of the Earth-Moon system and the internal pressures in Earth would have been close to the present-day. Figure~\ref{fig:press_evo_change}D-F shows the pressure increase in the interior of an Earth-like planet due to a reduction in its AM from a given initial value to the AM of Earth when the lunar spin axis underwent the Cassini state transition ($0.417~L_{\rm EM}$, black line). The pressure increase for magma-ocean bodies with the same composition, mass and initial AM of each of the post-impact bodies is also shown (points). 

The pressure increase during tidal recession after a high-AM Moon-forming impact would have been large. In general, the increase in CMB pressure is greater than about 30~GPa and can be as large as 70~GPa. As an example, consider a planet that has an AM of 3.0~$L_{\rm EM}$ before tidal recession (point a in Figure~\ref{fig:press_evo_change}E). As the Moon tidally recedes, the AM of the planet is reduced (red arrow in Figure~\ref{fig:press_evo_change}E) eventually reaching 0.417~$L_{\rm EM}$ at the Cassini state transition (point b in Figure~\ref{fig:press_evo_change}E). During this process the pressure at the CMB is increased from 75~GPa to 135~GPa, an increase of 60~GPa (double headed arrow in Figure~\ref{fig:press_evo_change}E). The pressure increase during tidal recession after the canonical impact would have been small, only a few gigapascals at the CMB.

\subsection{Internal pressures throughout accretion}
\label{sec:results:acc}

So far, we have only considered the effect of giant impacts on the internal pressures in Earth-like bodies. However, changes in thermal and rotational states driven by impacts affect the pressures in proto-planets of all sizes. It is beyond the scope of this work to calculate impacts between smaller bodies, but we can place a constraint on the pressures in less-massive post-impact bodies by considering the effect of AM on internal pressures.

It is expected that, due to the AM imparted by single or multiple giant impacts, most terrestrial planets rotate rapidly for much of accretion. The best investigation to date of the spin state of rocky planets during the giant-impact phase of accretion has been conducted by \cite{Kokubo2010}. They used an {\it N}-body simulation of planet formation with bimodal impact outcomes, either perfect merging or hit-and-run, and tracked the AM of each of the bodies in the simulation. The gray points with bars in Figure~\ref{fig:KG10_pressures}A and B show the mean and 1-$\sigma$ distribution of angular velocity and AM of the final bodies produced by the simulations in \cite{Kokubo2010}. In calculating the rotation rate, \cite{Kokubo2010} assumed that the bodies were spherical with a uniform density of 3000~kg~m$^{-3}$. In neglecting deformation of bodies due to rotation, \cite{Kokubo2010} overestimated the rotation rate of the bodies and so it is better to consider the AM of their final bodies. For comparison, the lines in A and B show three different estimates for the CoRoL. The gray line gives the critical angular velocity for breakup, the angular velocity at which the equator has a velocity equal to that of a Keplerian orbit, for the rigid, spherical bodies used by \cite{Kokubo2010}. The black line is the CoRoL for magma-ocean planets calculated using the HERCULES code (Section~\ref{sup:sec:methods:HERCULES}). The CoRoL for fluid bodies calculated using HERCULES differs from the classical critical spin limit for rigid, incompressible bodies, as the calculation includes the effects of changes in shape with increasing AM, the compressibility of real materials, and the variation in density with thermal state. The red line shows the CoRoL for a stratified, partially-vaporized body with a thermal profile typical of bodies after giant impacts calculated using HERCULES.

The distribution of angular momenta of bodies at the end of accretion found by \cite{Kokubo2010} demonstrates the high AM of bodies expected during accretion. Most of the bodies in the $N$-body simulations would exceed the CoRoL after a typical giant impact (red line) and form synestias. Upon cooling to magma-ocean planets most of the bodies would fall below the CoRoL but would still be rotating rapidly with periods of less than a few hours.

The high AM of bodies during accretion lowers their internal pressures. We can place bounds on the internal pressures during accretion, and specifically after giant impacts, by calculating the pressure in magma-ocean planets using HERCULES. Figure~\ref{fig:KG10_pressures}C shows the CMB pressure in Earth-like magma-ocean planets of varying AM (colored lines) and mass. The black line shows the CMB pressure at the CoRoL for magma-ocean planets. The range of internal pressures in rotating planets of all sizes is wide and CMB pressures can be dramatically lower than that for slowly rotating bodies (typically by up to a factor of two). For the Earth-mass post-impact bodies considered above, the pressures in condensed planets of given AM are typically an upper bound on the pressures immediately following the impact that produced them (Figure~\ref{fig:press_evo_abs}), and we expect this also to be the case for less massive bodies. Therefore, given the high-AM expected for proto-planets of all sizes during accretion, the internal pressures immediately after and between giant impacts are lower than those in condensed, non-rotating bodies throughout the giant-impact phase. The average internal pressures in terrestrial bodies during accretion are hence lower than has been previously assumed.

Fully quantifying the AM of planets during accretion, and hence their internal pressures, is challenging because partitioning of AM between the growing bodies, ejecta, and a planetesimal population is poorly understood. The AM of bodies during accretion was likely overestimated by \citep{Kokubo2010} as they neglected AM loss due to ejecta and the formation and evolution of satellites, but the magnitude of these effects is unknown. Figure~\ref{fig:KG10_pressures} shows that a non-negligible fraction of bodies in the simulations of \cite{Kokubo2010} are above the CoRoL for magma-ocean planets. Such bodies would be above the CoRoL even after condensation and would certainly have formed satellites upon cooling, reducing the AM of the central body. The calculation is also complicated by the fact that impacts from smaller, planetesimal sized bodies ($\leq 0.1 M_{\rm Earth}$), can also substantially perturb the AM of terrestrial planets. Reference \cite{Rufu2017} found that impacts onto Earth-mass planets can result in both AM increases and decreases of up to 1~$L_{\rm EM}$ and that multiple smaller impacts can impart substantial AM, following a random walk in AM. A wide range of impacts may therefore alter the internal pressures of terrestrial planets during accretion. Further work is needed to quantify the rotation rate of planets during formation in order to better determine the time evolution of internal pressures. 

\begin{figure}
\centering
\includegraphics{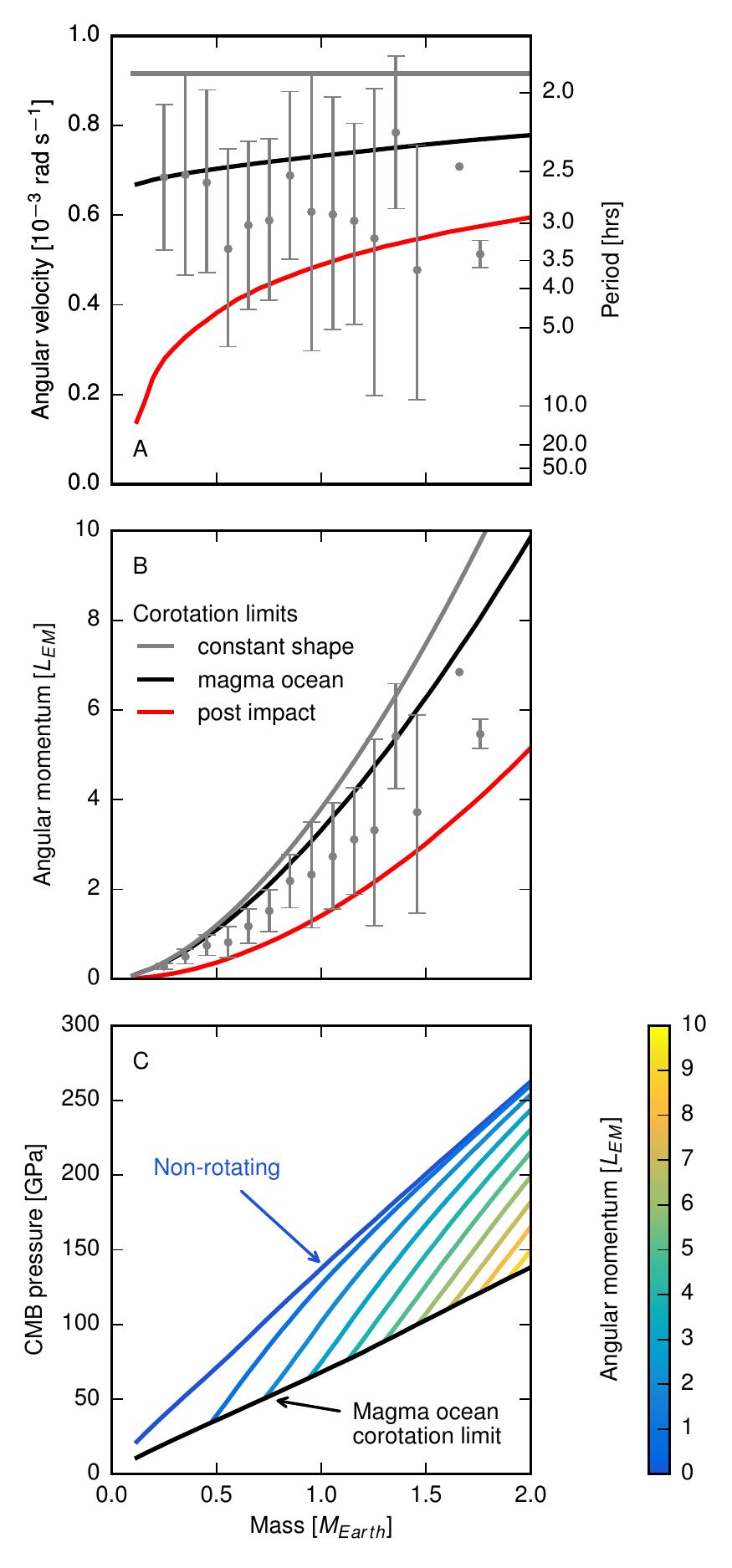}
\caption{Caption on next page.}
\end{figure}
\setcounter{figure}{5}
\begin{figure}
\caption{The rotation rates expected during the giant impact stage of planet formation could substantially reduce the internal pressure in bodies of a wide range of masses. (A) The angular velocity of different mass bodies at the end of accretion as calculated by \cite{Kokubo2010} (gray points and bars). The bars are the 1$\sigma$ standard deviation in the rotation rates found in their simulations. The gray line gives the critical angular velocity for breakup of a rigid, spherical body of bulk density 3000~kg~m$^{-3}$ as used by \cite{Kokubo2010}. The black and red lines give the corotation limit calculated using HERCULES for magma-ocean planets and substantially-vaporized, thermally-stratified bodies, respectively. (B) Angular momentum of different mass bodies as calculated by \cite{Kokubo2010} (gray points and bars). Notations and lines are the same as in A but in angular momentum space. (C) The pressure at the core-mantle boundary for Earth-like, magma-ocean planets of different masses and angular momenta in increments of 1~$L_{\rm EM}$ (colors) calculated using HERCULES. The black line gives the pressure at the corotation limit for a magma-ocean planet.}
\label{fig:KG10_pressures} 
\end{figure}

\section{Discussion}
\label{sec:discussion}

\subsection{Stochastic variation in pressure during accretion}
\label{sec:discussion:stoch}

It has been assumed that the pressures in terrestrial bodies depend only on mass and increase monotonically during accretion. However, we have shown that the internal pressures in bodies after giant impacts can be much lower than in condensed, non-rotating bodies (Figures~\ref{fig:press_evo_abs}A-C and \ref{fig:KG10_pressures}) and often lower than in the target before the impact (Figure~\ref{fig:target_press}). We have also demonstrated that impacts are followed by a period of increasing pressure during condensation and, in cases where a satellite is formed, tidal evolution (Figure~\ref{fig:press_evo_change}). 

During accretion, most planets experience several impacts with sufficient energy to vaporize a substantial fraction of the mantle \cite{Lock2017}. Furthermore, planets are expected to acquire significant AM due to either single of multiple impacts \cite{Kokubo2010}. Each impact that substantially changes the mass, thermal state or AM of a body alters its internal pressures. The magnitude and sign of the pressure change is highly sensitive to the parameters of the impact and is different for every collision. Furthermore, each impact is followed by a period of increasing pressure as the silicate vapor condenses, succeeded by a longer term decrease in pressure due to the tidal recession of satellites. The internal pressures in a body do not increase monotonically as it grows in mass but instead change stochastically in response to each impact. As terrestrial bodies are expected to rapidly rotate for much of accretion even between giant impacts, the average pressures are significantly lower than previously assumed.

Stochastic variation is a new paradigm for the evolution of internal pressures with implications for the physical and geochemical properties of terrestrial planets.

\subsection{Conditions for metal-silicate equilibration}
\label{sec:discussion:MSEs}

Core formation is a defining process in planet formation that controls the first order distribution of elements in terrestrial planets. The separation of elements between different reservoirs also allows for the use of radioactive systems (such as the Hf-W system, e.g., \cite{Jacobsen2005}) as chronometers of accretion, which are vital for constraining the time scales of planet formation. Partitioning of elements between metal and silicate is strongly dependent on the pressure and temperature of equilibration \cite[e.g.,][]{Li1996,Righter1997} and determining the conditions under which the cores of bodies form\ is fundamental to our understanding of planet formation.

The metal-silicate partitioning of MSEs is particularly sensitive to the conditions of equilibration. The concentrations of MSEs in Earth's mantle have been used to infer the pressure of equilibration during core formation \cite[e.g.,][]{Li1996,Righter1997} and hence the dynamics and chemistry of Earth's accretion \cite[e.g.,][]{Piet2017}. Assuming equilibration near the peridotite liquidus, studies have found that pressures in the range of 25 to 90~GPa near the end of Earth's accretion are required to reproduce the observed concentrations of MSEs \cite[e.g.,][]{Li1996,Righter1997,Yu2011a,Rubie2015a,Piet2017,Fischer2017}.

Metal-silicate equilibration and segregation of iron to the core occurs during and after giant impacts when the mantle of the post-impact body is substantially molten \cite{Rubie2007}. Despite the hot thermal state of bodies after giant impacts, metal is insoluble in silicate over much of the mantle and any free metal in the mantle falls under gravity towards the core. The gravitational settling time for iron in a molten mantle is shorter than the cooling timescale of the post-impact body \cite{Lock2018LPSC,Stevenson1990}, and the pressure and temperature profiles in the body immediately after the impact and during the first stages of evolution control the conditions of equilibration. Metal brought in by smaller impacts when the planet is solid between giant impacts is likely widely dispersed and held in the mantle until the next substantial whole-mantle melting event \cite{Stevenson1990}. Also, fractions of the metal from the core of the impactor and target would be dispersed in the molten mantle and/or quickly penetrate to the core with an as yet unknown degree of equilibration \cite{Canup2004,Cuk2012,Canup2012}. A common model has been that, after a giant impact, metal equilibrates in metal `ponds' at the bottom of the molten mantle atop a lower, mostly-solid silicate layer, before rapidly being transported to the core with little further equilibration \cite{Stevenson1990}. Based on calculations of the pressures in non-rotating, condensed bodies, the MSE pressure constraint has been interpreted as requiring that metal equilibrated at mid-mantle pressures at the bottom of a partial-mantle magma ocean. However, models of Moon-forming giant impacts suggest that most of the mantle would have been molten \cite{Nakajima2015}, a result that is expected to extend to all high-energy impacts. Any ponding of metal would have been deep in the body, above the pressure inferred from the MSE constraints. Furthermore, any significant solid layer above the CMB could have formed an effective barrier to segregation of metal to the core (see discussion in \cite{Stevenson1990}). A number of solutions have been proposed to solve this apparent contradiction, including that metal falling through the magma ocean did not equilibrate fully at each depth \cite{Rubie2007} or re-equilibration of the metal at each depth averaged to a lower mean equilibration pressure \cite{Li1996}, but these models have not been demonstrated to be physically realistic.

We have shown that the pressure in the mantle of terrestrial planets varies stochastically during accretion as a result of giant impacts. The average internal pressures, and hence the average pressure of metal-silicate equilibration, would be lower than if bodies were condensed and non-rotating for the whole of accretion. Lower internal pressures offer a mechanism to reconcile the expected degree of melting due to giant impacts with the geochemical evidence for metal-silicate equilibration at mid-mantle pressures. In particular, low mantle pressures are a predicted outcome of a high-AM lunar origin models. Equilibration near the CMB after high-AM Moon-forming impacts is consistent with the constraint from MSEs.

\subsection{Crystallization of the terrestrial magma ocean}
\label{sec:discussion:freezing}

As discussed in the introduction, the pressures in the interior of a planet control how the mantle freezes as the relative slopes of the liquid adiabat and liquidus change with pressure. Bodies with lower CMB pressures (less than around 70 to 105~GPa \cite{Stixrude2009,Thomas2013}) freeze from the bottom of the mantle upwards whereas those with higher CMB pressures may freeze from the middle outwards. If Earth after the Moon-forming impact had close to its present-day pressure structure, such as in the canonical scenario, the CMB pressure would have been high enough for the mantle to have frozen from the middle outwards, isolating a basal magma ocean in the lower mantle \cite{Labrosse2007}. 

However, we have shown that the pressure in the terrestrial magma ocean after high-AM Moon-forming impacts would have been much lower (Figure~\ref{fig:press_evo_abs}D-F). The pressure at the CMB would have been either lower than, or coincident with, the proposed intersection of the adiabat and liquidus (Figure~\ref{fig:press_evo_abs}). The mantle would start to freeze quickly after the condensation of the silicate vapor while the internal pressures were low as the liquid upper mantle would have allowed efficient extraction of heat \cite{Lebrun2013,Zahnle2015}. The mantle would probably have frozen from the bottom up and there would have been no substantial basal magma ocean after a high-AM Moon-forming impact. 

At the CMB pressures expected after high-AM giant impacts, the multiphase adiabat and liquidus are close to parallel \cite{Stixrude2009,Thomas2013}. The density of the melt is also similar to that of the solid in this pressure range \cite{Thomas2013}. In the initial stages of magma ocean cooling, when the lower mantle was still crystallizing, the cooling timescale was likely shorter than the percolation timescale and a large fraction of the lower mantle could have frozen together in bulk. 

How the mantle froze is substantially different between the canonical and high-AM Moon formation models. Dynamical modeling that links magma ocean crystallization to the production of different chemical reservoirs could be used to test different lunar origin models.

\subsection{Melting of the lower mantle during lunar tidal recession}
\label{sec:discussion:pressure_melting}

We have shown that formation of a substantial basal magma ocean after a high-AM Moon-forming impact is unlikely, but increases in the CMB pressure during tidal recession could still produce partial melt in the lowermost mantle. During tidal recession, parts of the mantle were either partially molten or close to the solidus. For example, although the lowermost mantle likely froze quickly after condensation of the silicate-vapor atmosphere \cite[e.g.,][]{Lebrun2013}, the thermal boundary layer at the CMB would have remained hot enough for silicates to be partially molten. The increase in pressure during tidal recession could cause material near the phase boundary to melt or solidify.

The effect of pressure changes on melting or freezing is dictated by, among other factors, the relative slopes of the liquidus, solidus, and the liquid and solid adiabats. The relative slopes are a topic of some debate but, for the purposes of our discussion, we will accept the proposal that the slopes of the liquidus and liquid adiabat cross around 70 to 105~GPa \cite{Stixrude2009,Thomas2013}. For simplicity, we will further assert that the slopes of the solidus and solid adiabat cross at the same pressure. In other words, we assume that both the solidus and liquidus have a maximum in entropy at the same pressure. Future work will explore the influence of alternative relationships between the phase boundaries and adiabats on pressure induced melting/freezing.

At pressures lower than that at which the slopes of the phase boundaries and adiabats cross, an increase in pressure would promote freezing. At higher pressures, increasing pressure during tidal recession would have driven partial melting. The pressure increase during tidal recession could therefore have produced a substantial mass of melt in the lower mantle. If the melt was denser than the solid \cite{Thomas2013}, percolation could have separated the melt from the solid, potentially collecting a layer of melt above the CMB.

A major focus of deep Earth research is the study of the properties, origin and evolution of seismically anomalous regions observed in the lower mantle, known as large low-shear-velocity provinces (LLSVPs) \cite[e.g.,][]{Trampert2004,Ishii1999} and ultra low velocity zones (ULVZs) \cite[e.g.,][]{Garnero1995}. These features could be explained by the presence of partial melt \cite[e.g.,][]{Williams1996} or compositional heterogeneity \cite[e.g.,][]{Wicks2017}. Most mantle plumes appear to be sourced from these regions \cite[e.g.,][]{Austermann2014} leading to the development of the idea that they may be the source of isotopic anomalies measured in ocean island basalts and large igneous provinces \cite[e.g.,][]{Mukhopadhyay2012,Rizo2016}. Some of these isotopic signatures are ancient, having formed within tens of million years of the start of the solar system \cite{Mukhopadhyay2012,Rizo2016}. If the observed lower mantle heterogeneities are associated with these isotopic anomalies, they likely formed early in Earth's history and survived to the present day. How ancient heterogeneities could have formed and then preserved until the present day is a matter of ongoing debate. One proposed formation mechanism is that fractional crystallization of a basal magma ocean would have left iron-enriched, dense material at the bottom of the mantle \citep{Labrosse2007}. A solid, or still partially molten, iron-enriched layer may be consistent with the observed seismic anomalies \citep[e.g.,][]{Wicks2017,Williams1996}.

Pressure-induced melting offers an alternative mechanism to create lower-mantle heterogeneity early in Earth's history. If the melt produced was fully or partially separated from the solid residue, pressure-induced melting would have produced incompatible-element-enriched regions in the lower mantle. Such heterogeneity, or even just the persistence of partial melt, could potentially explain the seismically anomalous regions observed in the lower mantle today.

\section{Conclusions}
\label{sec:conclusions}

Until now it has been assumed that the pressures in terrestrial planets depended only on their mass and increased monotonically during accretion. We have shown that this assumption does not hold and that knowledge of the thermal and rotational state of bodies is needed to infer their internal pressures. The pressures in terrestrial planets change stochastically through accretion as a result of high-energy and/or high-AM impacts. In particular, the pressure in Earth after a high-AM Moon-forming impact would be much lower than at the present day. 

Stochastic change is a new paradigm for the evolution of internal pressures during accretion. The resulting lower average pressures help reconcile the observed concentrations of moderately siderophile elements in the terrestrial mantle with the expected degree of melting and metal-silicate partitioning after giant impacts. Equilibration at the core-mantle boundary pressure after a high-AM Moon-forming impact would provide the modest equilibration pressure inferred from the observations \cite[e.g.,][]{Li1996,Righter1997,Yu2011a,Rubie2015a,Piet2017,Fischer2017}. Differences in the internal pressures after different Moon-formation scenarios change how the mantle solidifies. There would be no substantial basal magma ocean after high-AM Moon-forming impacts. Difference in internal pressures and pressure evolution opens pathways to use geochemical and geophysical observations of the present-day Earth to test different Moon-formation scenarios. 

In addition, we show that cooling and tidal evolution of satellites after giant impacts can lead to increases in pressure. Pressure increases in the lower mantle could lead to partial melting. Pressure-induced melting provides a new mechanism for the creation of chemical heterogeneity early in Earth's history and could explain the existence of seismologically anomalous regions in the present-day lower mantle \cite{Trampert2004,Ishii1999}. Pressure-induced phase transitions during the recovery of a body after a giant impact are a previously unrecognized phenomenon in planet formation.


\renewcommand{\thesection}{A\arabic{section}}   
\renewcommand{\thetable}{A\arabic{table}}   
\renewcommand{\thefigure}{A\arabic{figure}}
\renewcommand{\theequation}{A\arabic{equation}}

\setcounter{section}{0}
\setcounter{figure}{0}
\setcounter{table}{0}
\setcounter{equation}{0}

\section{Materials and methods}
\label{sup:sec:methods}

\subsection{Smoothed particle hydrodynamics}
\label{sup:sec:methods:SPH}

To demonstrate the range of pressures in post-impact states, we analyzed the post-impact bodies produced by smoothed particle hydrodynamic (SPH) giant impact simulations. We combined the simulations presented in \cite{Lock2017} with new impact calculations (Table~\ref{sup:tab:impacts}). These new simulations include more impacts near the escape velocity with large projectile to target mass ratios. We modeled giant impacts using the GADGET-2 SPH code \cite{Springel2005} modified for planetary impact studies \cite{Marcusthesis}, which has been used in several giant impact studies \cite[e.g.,][]{Lock2017,Cuk2012,Rufu2017}. The colliding bodies were differentiated (2/3 rocky mantle, 1/3 iron core by mass) with forsterite mantles and iron cores modeled using M-ANEOS equations of state \cite{Canup2012,Melosh2007}. See \cite{Cuk2012} or \cite{Lock2017} for a full description of the methods. We simulated collisions with a large range of impact parameters that lead to approximately Earth-mass final bodies. The resulting post-impact structures have a wide range of thermodynamic and rotational states (Table~\ref{sup:tab:impacts}).

To quantify the energy of each impact, we use a modified specific energy, $Q_{\rm S}$ \cite{Lock2017}. $Q_{\rm S}$ is defined as 
\begin{linenomath*}
\begin{equation}
Q_{\rm S} = Q'_{\rm R} \left ( 1 + \frac{M_{\rm p}}{M_{\rm t}} \right ) (1-b) \;\;,
\label{sup:eqn:QS}
\end{equation}
\end{linenomath*}
where $Q'_{\rm R}$ is a center of mass specific impact energy modified to include only the interacting mass of the projectile \cite{Leinhardt2012}. Each of the terms in Equation \ref{sup:eqn:QS} accounts for a factor that affects how efficiently energy is coupled into the shock pressure field in the impacting bodies. \cite{Lock2017} showed that the  specific entropy of the mantles of post-impact bodies scales well with $Q_{\rm S}$, making it a good predictor of post-impact thermal state. 

The qualitative relationship of each post-impact body to the corotation limit (CoRoL) was ascertained visually in the same manner as in \cite{Lock2017}.

SPH self-consistently calculates the pressure of material in post-impact bodies, and calculating the pressure in roughly compositionally homogeneous regions of a body is straightforward. We determined the central pressure of bodies by averaging the pressure of the 50 highest-pressure SPH particles. We defined the pressures at 50~wt\% of the mantle as the pressure contour below which 50\% of the bound silicate mass resided.

Inferring the CMB pressure from SPH calculations is made difficult by the well known issue of resolving high density contrasts in SPH. The density contrast between core and mantle can lead to a layer of anomalous pressure particles on either side of the boundary (see e.g., Figure~3 in \cite{Lock2017}). Additionally, in post-impact states, the CMB can be somewhat blurred by the mixing of high-specific entropy iron particles into the lower mantle.  In order to compensate for these two issues, we calculated the mass of silicate and iron particles in the midplane of a structure in a series of 300~km wide radial bins. We defined the CMB as the boundary between the set of bins which were majority iron by mass and those that were majority silicate. We took the CMB pressure as the mass-weighted average of the SPH particles in the two bins either side of the CMB. This method can be shown to accurately calculate the CMB pressure (Section~\ref{sup:sec:methods:comparison}).

\subsection{HERCULES}
\label{sup:sec:methods:HERCULES}

We calculate the structure of corotating bodies using the newly developed HERCULES code \cite{Lock2017}.  HERCULES utilizes a potential field method to calculate the equilibrium structure of planets with a given thermal state, composition, mass and AM, using realistic equations of state. 

In HERCULES, a body is described as a series of nested concentric spheroids. All the material between the surfaces of any two consecutive spheroids is called a layer. The internal pressures in HERCULES are calculated by a first-order integration of the combined gravitational and centrifugal potential in the equator \cite{Lock2017}. The central and CMB pressures are naturally derived using this formulation. The pressure at 50~wt\% of the mantle was linearly interpolated between the mass fractions contained within the equipotential surfaces bounding the desired mass fraction.  

To allow direct comparison to our SPH post-impact bodies, we calculated the structure of bodies using the same equations of state as used in the impact simulations. For magma-ocean planets we used isentropic cores and mantles with specific entropies of 1.5 and 4~kJ~K$^{-1}$~kg$^{-1}$ respectively. This core isentrope has a temperature of 3800~K at the pressure of the present-day CMB, similar to the present thermal state of Earth's core (see Figure~\ref{sup:fig:SpT} for the pressure-temperature profiles for example forsterite isentropes). The mantle isentrope intersects the liquid-vapor phase boundary at low pressure (10~bar) and about 4000~K. We used a surface bounding pressure of 10~bar so as not to resolve the structure of the silicate vapor atmosphere. Our chosen thermal state approximates that of a well-mixed, mostly-liquid, magma-ocean planet.

The relative timings of the freezing of the mantle and tidal recession of the Moon is uncertain and so here we have used a magma-ocean thermal profile for all condensed planets. However, the thermal state chosen for the magma-ocean planet has little effect on the internal pressures. Figure~\ref{sup:fig:HERCULES_therm} shows the absolute and fractional difference in internal pressures in bodies calculated using mantle isentropes of 3, 3.2 and 4~kJ~K$^{-1}$~kg$^{-1}$. Mantle specific entropies of 3 and 3.2~kJ~K$^{-1}$~kg$^{-1}$ correspond to mantle potential temperatures of $\sim 1600$~K and $\sim 1900$~K, similar to the present-day and early terrestrial mantle respectively. The pressure in the lower entropy bodies are higher by a few percent, e.g., up to 4~GPa at the CMB, which is not significant for the conclusions of this work. The difference in the pressure change during tidal evolution for each of these bodies is even smaller.

The partially-vaporized, thermally-stratified planets used to calculate the properties of bodies just below the CoRoL in Figures~\ref{fig:KG10_pressures} and \ref{sup:fig:pCoRoL} were defined as having the same thermal profile as a magma-ocean planet, but with a hotter upper mantle (25~wt\% of the mantle). The upper mantle was isentropic with a specific entropy of $S_{\rm outer}$ until the thermal profile intersected the liquid-vapor phase boundary. At pressures below the intersection the body was assumed to be pure vapor on the liquid-vapor phase boundary. This structure was called a stratified structure in \cite{Lock2017} and approximates the substantially vaporized and stratified thermal structure produced by giant impacts. 

We did not directly calculate the structure for all the bodies in this paper except when calculating the properties of Earth-mass planets. Instead, we calculated grids of HERCULES planets with a range of properties. For magma-ocean planets we varied the mass, core-mass fraction, and AM. We used total mass increments of 0.1~$M_{\rm Earth}$, core-mass fraction increments of 0.05, and a base AM increment of 0.1~$L_{\rm EM}$ for bodies with masses $>0.2$~$M_{\rm Earth}$. For 0.1~$M_{\rm Earth}$ and 0.2~$M_{\rm Earth}$ bodies, we used base AM increments of 0.02 and 0.05~$L_{\rm EM}$ respectively. When HERCULES failed to converge at the next AM step, higher AM runs were performed using smaller AM steps. The AM step was sequentially halved five times to provide finer AM resolution just below the CoRoL. The internal pressures at different layers were calculated for each of these bodies and linear interpolation used to calculate the properties for a body of a given composition, mass and AM. The variation in internal pressures in the mass, composition and AM space we consider in this paper is close to linear and this technique gives a very good approximation to the pressures in directly calculated HERCULES planets. 

To find the properties of partially-vaporized, thermally-stratified bodies at the CoRoL we again interpolated a grid of HERCULES planets with varying mass, core-mass fraction, and $S_{\rm outer}$. We ran the same mass and core-mass fraction increments as for the magma-ocean planets and $S_{\rm outer}= 4$, 4.25, 4.5, 4.75, 5.0, 5.5, 6.0, 6.5, 7.0, 7.5, and 8.0~kJ~K$^{-1}$~kg$^{-1}$. For each parameter set, the structures of bodies were calculated with the same AM step procedure. We found the AM of the CoRoL for each set of parameters as described in \cite{Lock2017}. The internal pressures in a body at the CoRoL were found by linearly extrapolating from the calculated corotating planets in the same manner as in \cite{Lock2017} (see their Figure~S8). For the CoRoL shown in Figure~\ref{fig:KG10_pressures}, a fixed outer specific entropy of $S_{\rm outer} =6$~kJ~K$^{-1}$~kg$^{-1}$ was used, which is typical of post-impact bodies \cite{Lock2017}. This profile intersects the liquid-vapor phase boundary at about 10~kbar. For a post-impact body with a given mass, core-mass fraction and AM, the properties of the body when it has cooled to just below the CoRoL shown in Figure~\ref{sup:fig:pCoRoL} were found by linearly interpolating this grid of CoRoL properties in mass-core fraction-AM space. 

For this paper, we used the same HERCULES parameters as used in \cite{Lock2017}. The internal pressures calculated using HERCULES are only weakly dependent on the number of concentric potential layers (Figure~\ref{sup:fig:HERCULES_Nlay}), the number of points used to describe potential surfaces (Figure~\ref{sup:fig:HERCULES_Nmu}), and the maximum spherical harmonic degree included in the calculation (Figure~\ref{sup:fig:HERCULES_kmax}). For the range of parameters we considered, the internal pressures varied by less than a 0.5\%. 

\subsection{Comparison of methods}
\label{sup:sec:methods:comparison}

The physical structures produced by SPH and HERCULES for corotating bodies with a range of thermal states are in good agreement \cite{Lock2017}. Here we examine in more depth the internal pressures calculated using our two methods. 

Figure~\ref{sup:fig:HERCULES_comparison} shows a comparison of the pressures calculated using SPH and HERCULES for Earth-mass, corotating bodies with isentropic mantles of varying specific entropy. The shape and pressure contours for these bodies are shown in Figure~4 of \cite{Lock2017}. The central pressures calculated using the two methods are in agreement to within a few percent. On average, the calculated CMB pressures show similar levels of agreement, but there is a larger variance between different bodies. This is not surprising given the necessity of averaging the pressure around the CMB; nevertheless the agreement is generally better than 10~GPa. The fractional difference in the mid-mantle pressure is larger than for the center or CMB but the absolute error is only on the order of a few gigapascals. Given the agreement between the two different methods we are confident in the conclusions of this work.

\section*{Supplementary materials}
\noindent Figure~\ref{sup:fig:pCoRoL}. Pressure change in cooling from the CoRoL to a magma-ocean planet.
\\\noindent Figure~\ref{sup:fig:pchange_cooling_Moon}. Effect of forming the Moon on internal pressures.
\\\noindent Figure~\ref{sup:fig:SpT}. Isentropes for the M-ANEOS derived forsterite EOS in pressure-temperature space.
\\\noindent Figure~\ref{sup:fig:HERCULES_therm}. Effect of thermal state on the pressure in condensed bodies.
\\\noindent Figure~\ref{sup:fig:HERCULES_Nlay}. Sensitivity to the number of concentric layers using in HERCULES.
\\\noindent Figure~\ref{sup:fig:HERCULES_Nmu}. Sensitivity to the number of points used to describe each surface in HERCULES.
\\\noindent Figure~\ref{sup:fig:HERCULES_kmax}. Sensitivity to the maximum spherical harmonic degree used in HERCULES.
\\\noindent Figure~\ref{sup:fig:HERCULES_comparison}. Comparison of pressures calculated using SPH and HERCULES.
\\\noindent Table~\ref{sup:tab:impacts}. Impact parameters and properties of resulting bodies at different stages in evolution.


\bibliography{References}

\section*{Acknowledgements}
\noindent We are grateful to two anonymous reviewers for their comments that helped improve the quality of this manuscript. We would also like to thank Jennifer Jackson and Paul Asimow for useful discussions. {\bf Funding:} This work was supported by NESSF grant NNX13AO67H (SJL), NASA grant NNX15AH54G (STS), and DOE-NNSA grant DE-NA0002937 (STS). SJL also gratefully acknowledges support from Harvard University's Earth and Planetary Sciences Department and Caltech's Division of Geological and Planetary Sciences. 
{\bf Author Contributions:} STS performed the SPH calculations. SJL performed the HERCULES calculations, analyzed all simulation output and wrote the paper. Both authors discussed the results and contributed to the ideas in the paper.
{\bf Competing interests:} The authors declare that they have no competing interests. 
{\bf Data and materials availability:} All data needed to evaluate the conclusions in the paper are present in the paper and/or the Supplementary Materials. The modified version of GADGET-2 and the EOS tables are contained in the supplement of \cite{Cuk2012}. The HERCULES code is included in the supporting information of \cite{Lock2017} and is available through the GitHub repositry:  \url{https://github.com/sjl499/HERCULESv1\_user}. 



\renewcommand{\thepage}{S\arabic{page}}  
\renewcommand{\thesection}{S\arabic{section}}   
\renewcommand{\thetable}{S\arabic{table}}   
\renewcommand{\thefigure}{S\arabic{figure}}
\renewcommand{\theequation}{S\arabic{equation}}

\setcounter{page}{1}
\setcounter{section}{0}
\setcounter{figure}{0}
\setcounter{table}{0}
\setcounter{equation}{0}

\title{Supplementary materials for ``Giant impacts stochastically change the internal pressures in terrestrial planets"}

\authors{Simon J. Lock\affil{1,2}, Sarah T. Stewart\affil{3}}

\affiliation{1}{Division of Geological and Planetary Sciences, Caltech}
\affiliation{2}{Department of Earth and Planetary Sciences, Harvard University}
\affiliation{3}{Department of Earth and Planetary Sciences, U. California Davis}

\correspondingauthor{Simon J. Lock}{slock@caltech.edu}

\vspace{20px}
Supplementary materials include:
\begin{enumerate}
\item Figures S1-S8. 
\item Table S1. 
\end{enumerate}

\begin{figure}
\centering
\includegraphics{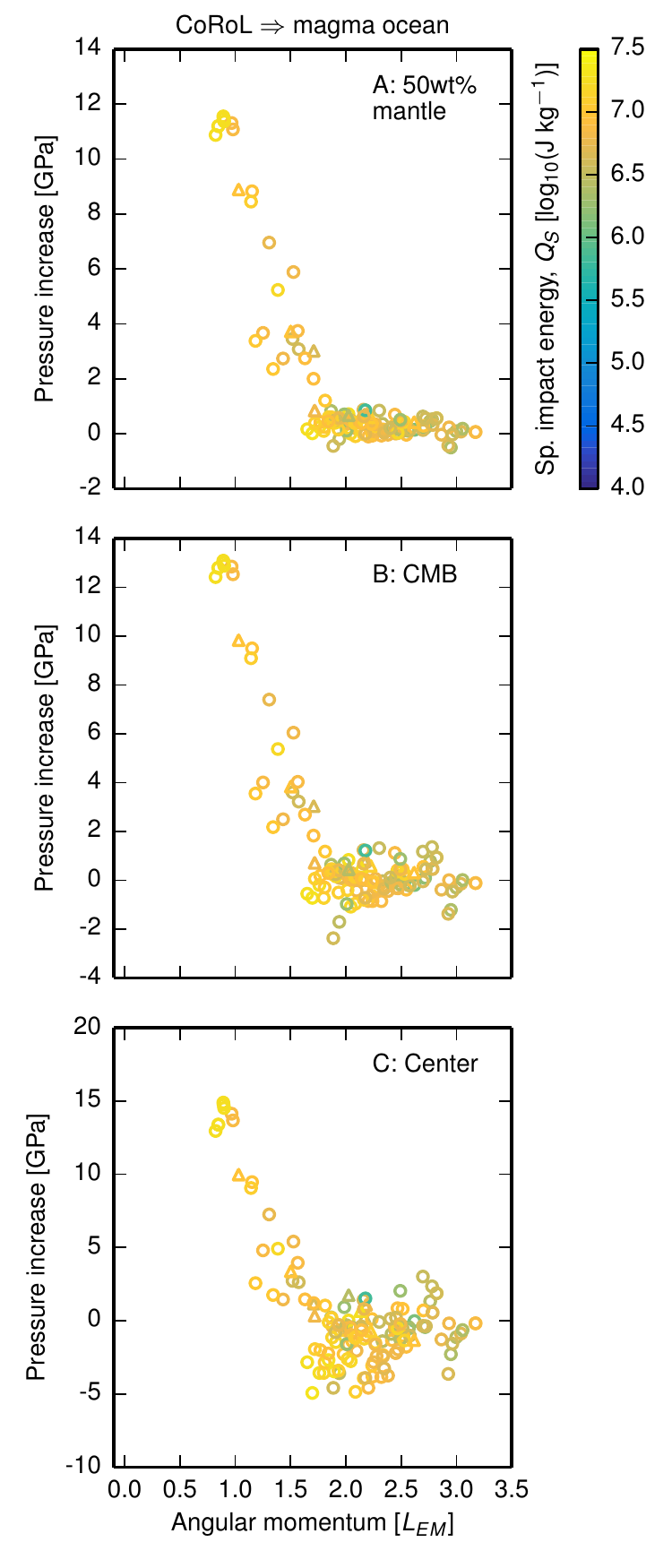}
\caption{The pressure change in cooling from the corotation limit to a magma ocean is small for high-angular-momentum bodies. Panels show the pressure difference between a body just below the CoRoL and a magma-ocean planet of the same angular momentum at the middle of the mantle by mass (A), the core-mantle boundary (B) and at the center (C) of the body. Colors and symbols are the same as in Figure~\ref{fig:press_evo_abs}. To calculate the pressures in bodies at the corotation limit, we used a thermal profile that approximates the thermally-stratified, partially-vaporized structure of post-impact bodies during cooling (see Section~\ref{sup:sec:methods}). The competing effects of changes in moment of inertia and bulk density during condensation means that some bodies have slightly higher internal pressures when they are just below the the CoRoL than when a magma-ocean planet.
}
\label{sup:fig:pCoRoL}
\end{figure}

\begin{figure*}
\centering
\includegraphics[width=\textwidth]{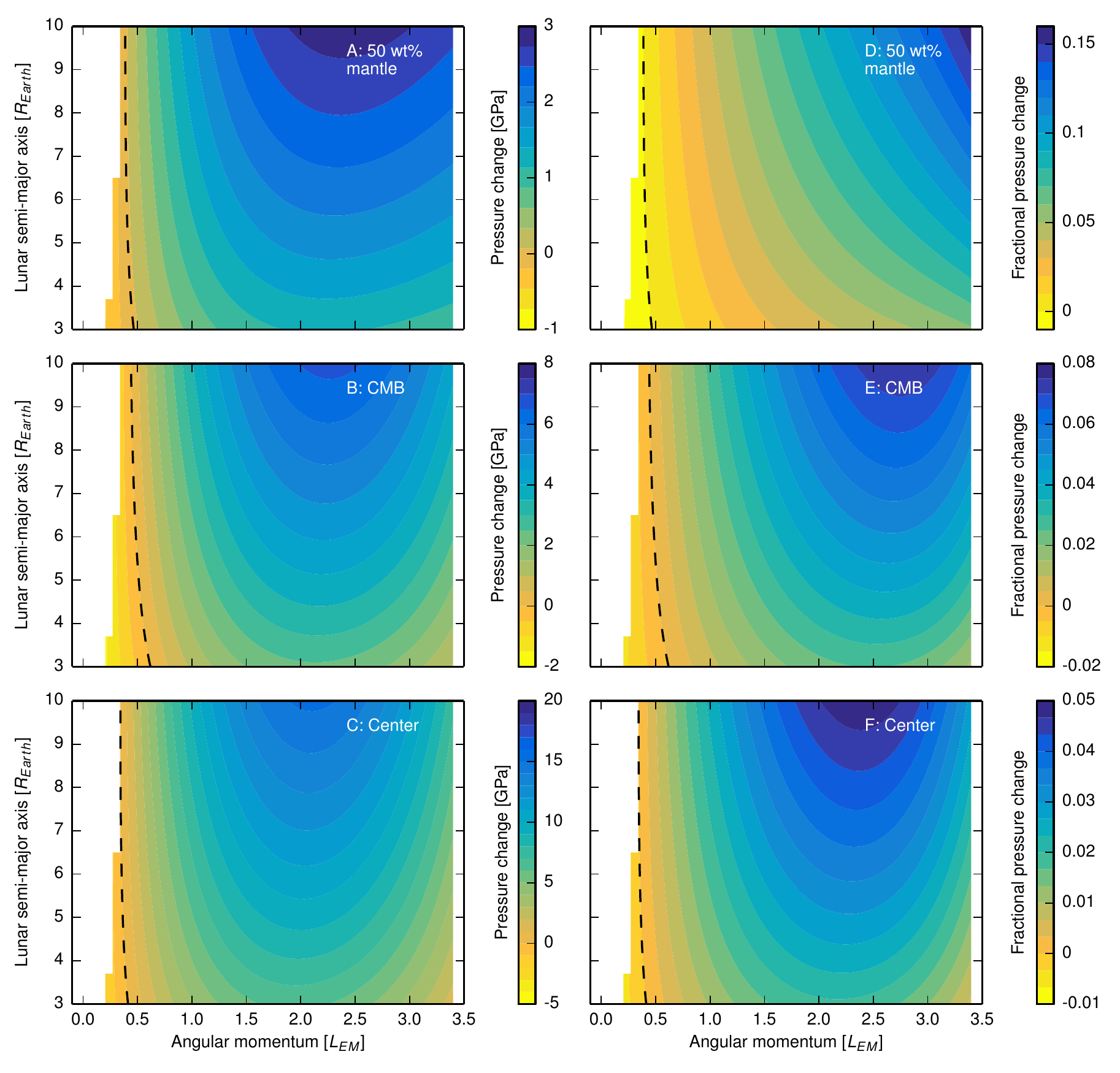}
\caption{The formation of a close-in moon has only a minor effect on the internal pressures in a magma-ocean after a giant impact. Shown is the absolute (left column) and fractional (right column) difference in the pressures in an Earth-like magma-ocean planet orbited by a tidally-locked Moon at a given semi-major axis (y-axis) and a system with the same total angular momentum, mass, and composition, but with all the mass combined into a single magma-ocean planet. Panels show the pressure difference at the middle of the mantle by mass (A, D), the core-mantle boundary (B, E) and at the center (C, F) of the body. The dashed lines indicate the locus of points for which there is no difference in pressure between the two systems.
}
\label{sup:fig:pchange_cooling_Moon}
\end{figure*}

\begin{figure}
\centering
\includegraphics{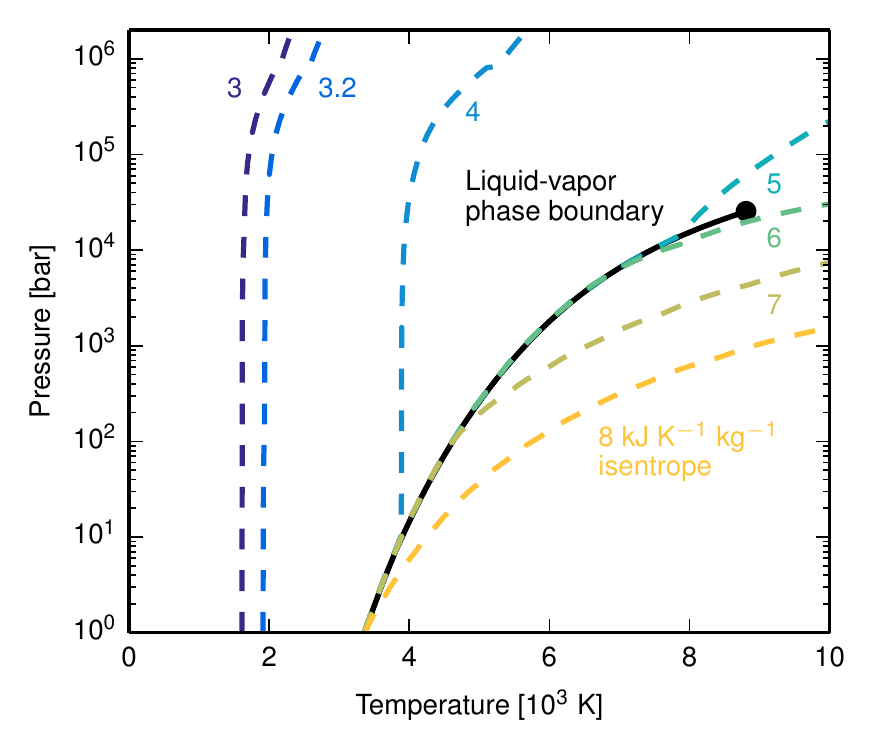}
\caption[]{Isentropes for the M-ANEOS derived forsterite EOS used in this work in pressure-temperature space. Each colored line is an isentrope for the specific entropy given by the number of the same color in kJ~K$^{-1}$~kg$^{-1}$. The black line is the liquid-vapor phase boundary. The black dot is the critical point. Adapted from \cite{Lock2017}.
}
\label{sup:fig:SpT}
\end{figure}


\begin{figure*}
\centering
\includegraphics{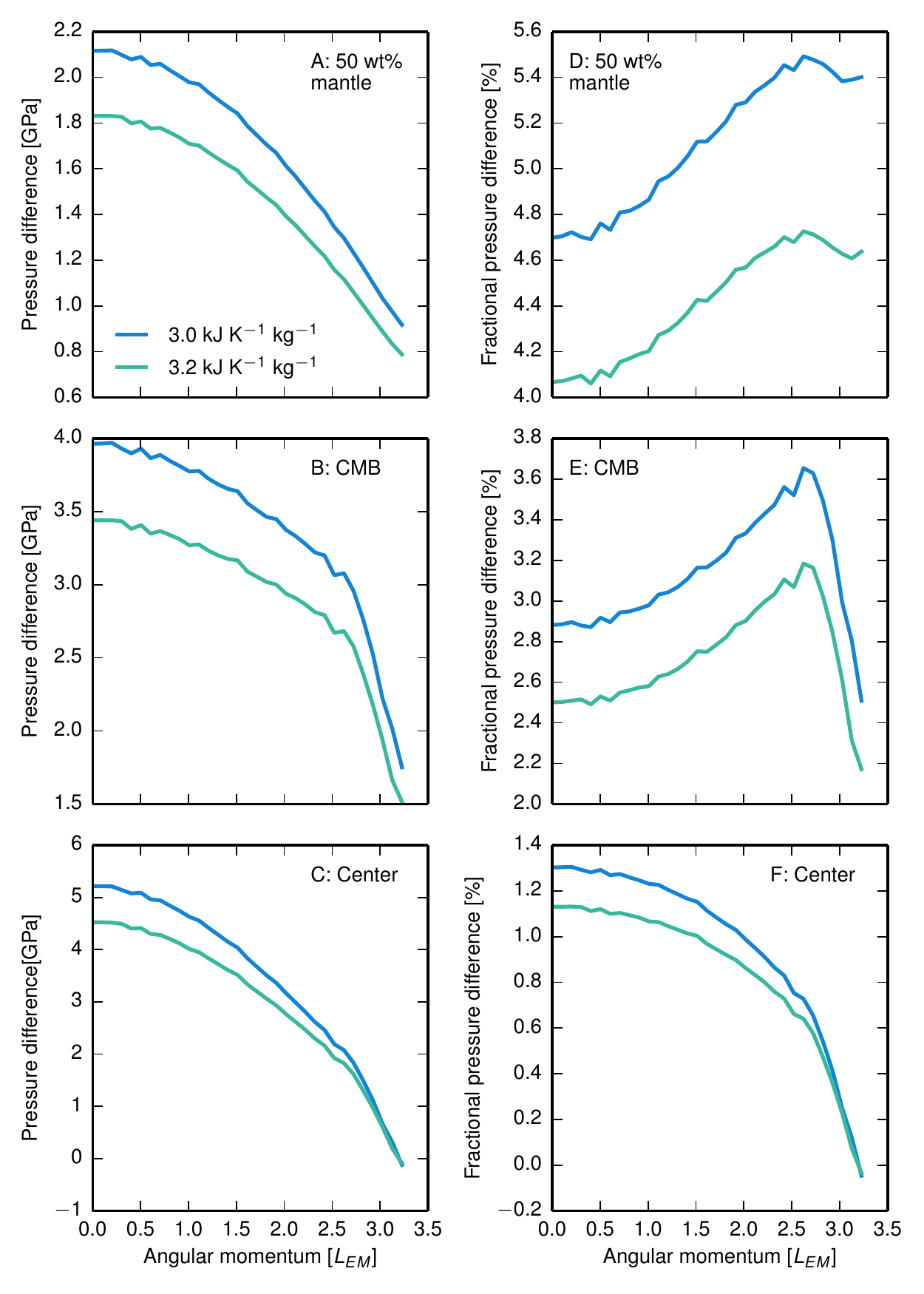}
\caption{The thermal state of condensed terrestrial bodies makes only a small difference to the internal pressure. The absolute (left column) and fractional (right column) difference in pressure between bodies that have a mantle entropy of 4~kJ~K$^{-1}$~kg$^{-1}$ (the thermal state used for magma-ocean planets elsewhere in this paper) and bodies with mantle entropies of 3 and 3.2~kJ~K$^{-1}$~kg$^{-1}$ are shown by the blue and green lines respectively. Mantle specific entropies of 3 and 3.2~kJ~K$^{-1}$~kg$^{-1}$ correspond to mantle potential temperatures of $\sim 1600$~K and $\sim 1900$~K, similar to the present-day and early terrestrial mantle respectively. Rows show the pressure difference in the mid mantle (A, D), core-mantle boundary (B, E), and center of the body (C, F). }
\label{sup:fig:HERCULES_therm}
\end{figure*}

\begin{figure}
\centering
\includegraphics{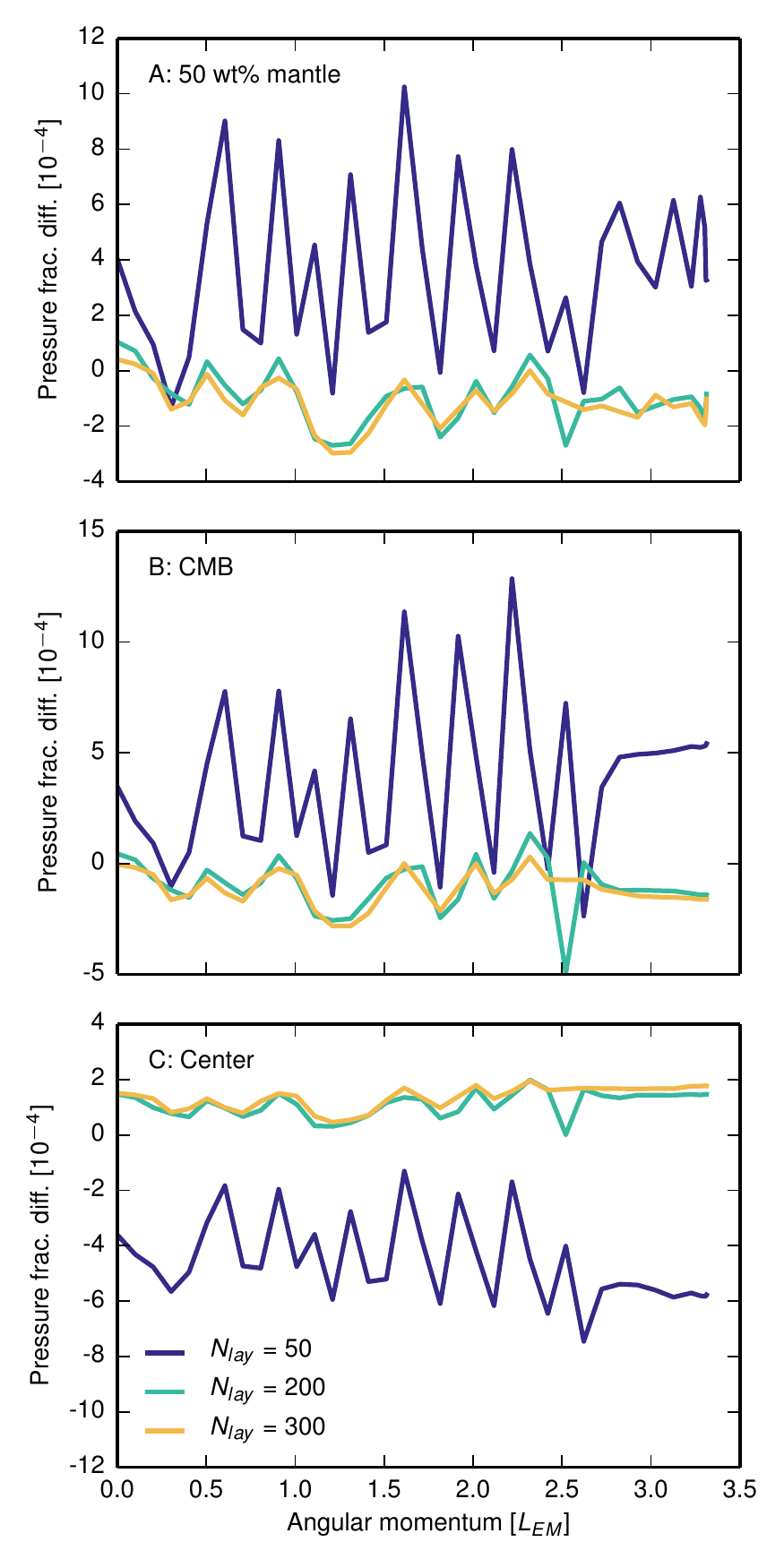}
\caption{The internal pressures calculated using HERCULES are only weakly dependent on the number of concentric layers used ($N_{\rm lay}$). Shown are the fractional difference in pressure in bodies with a range of angular momenta calculated using HERCULES with a given number of concentric layers ($N_{\rm lay}$, colored lines) and the same body calculated with $N_{\rm lay}=100$ as used elsewhere in this paper. Panels show the difference in the pressure at the middle of the mantle by mass (A), the core-mantle boundary (B) and the center (C) of the bodies. 
}
\label{sup:fig:HERCULES_Nlay}
\end{figure}

\begin{figure}
\centering
\includegraphics{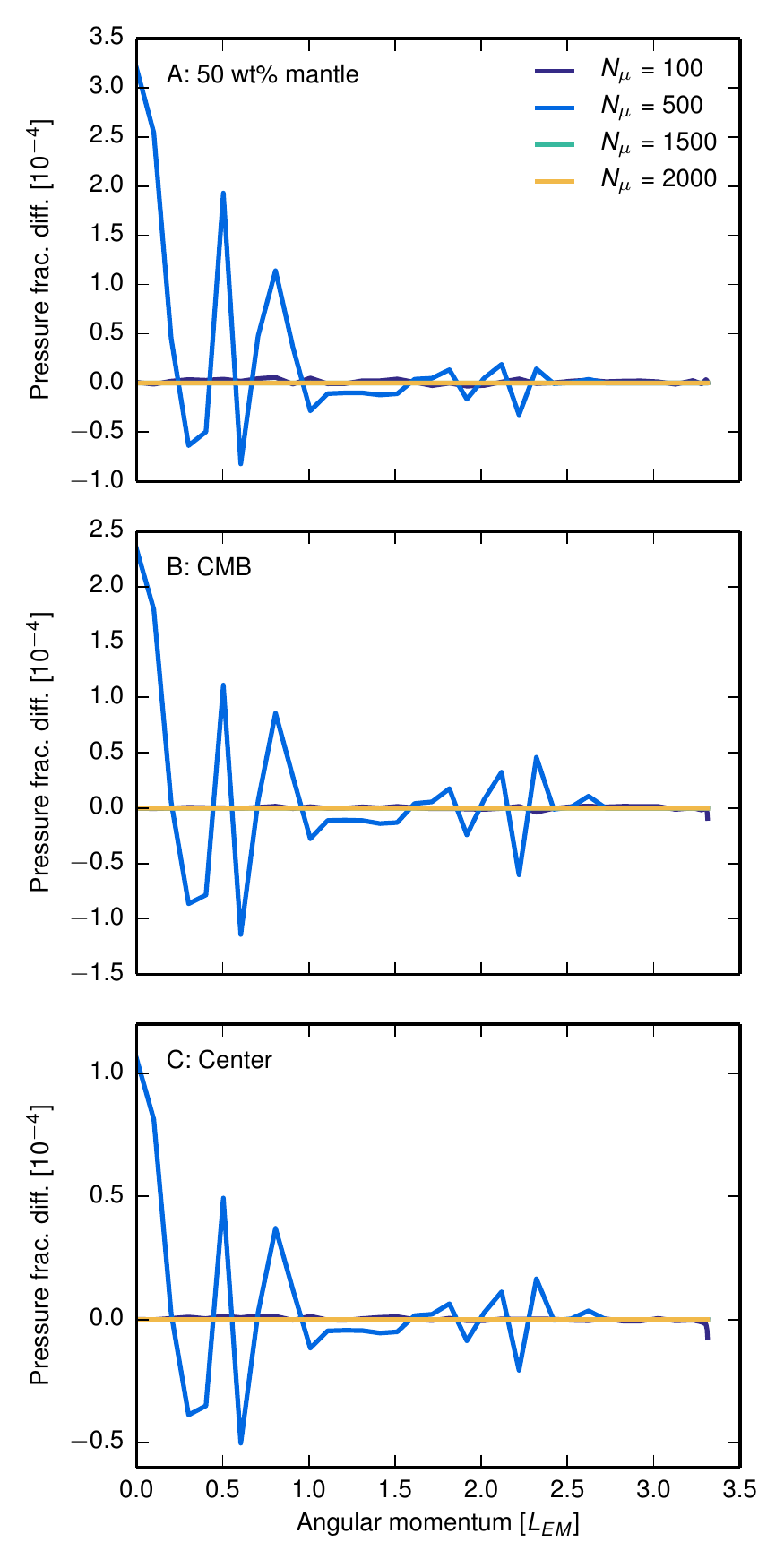}
\caption{The internal pressures calculated in HERCULES are only weakly dependent on the number of points used to describe each equipotential surface ($N_{\mu}$). Shown are the fractional differences in internal pressures between bodies with a range of angular momenta calculated using HERCULES with varying $N_{\mu}$ (colored lines) compared to the same body calculated using the parameters used elsewhere in this paper ($N_{\mu}=1000$). Panels show the difference in the pressure at the middle of the mantle by mass (A), the core-mantle boundary (B) and the center (C) of the bodies.
}
\label{sup:fig:HERCULES_Nmu}
\end{figure}

\begin{figure}
\centering
\includegraphics{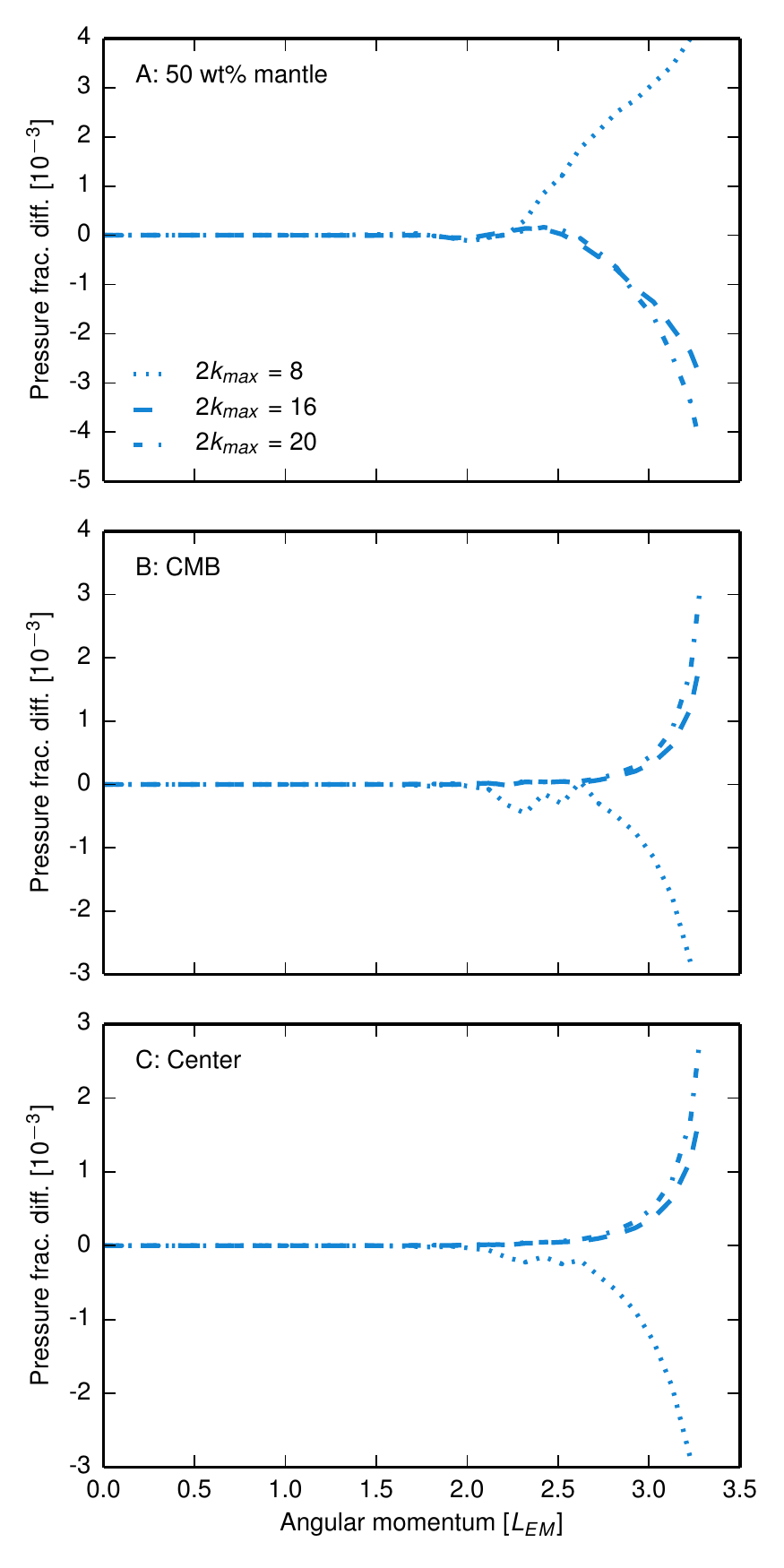}
\caption{The internal pressures calculated in HERCULES are only weakly dependent on the maximum spherical harmonic degree included ($2k_{\rm max}$). Shown are the fractional differences in internal pressures between bodies with a range of angular momenta calculated using HERCULES with different $k_{\rm max}$ (lines) compared to the same body calculated using the parameters used elsewhere in this paper ($k_{\rm max}=6$). Panels show the differences in the pressure at the center (A), the core-mantle boundary (B) and the middle of the mantle by mass (C) of each body. Panels show the difference in the pressure at the middle of the mantle by mass (A), the core-mantle boundary (B) and the center (C) of the bodies.
}
\label{sup:fig:HERCULES_kmax}
\end{figure}

\begin{sidewaysfigure*}
\centering
\includegraphics{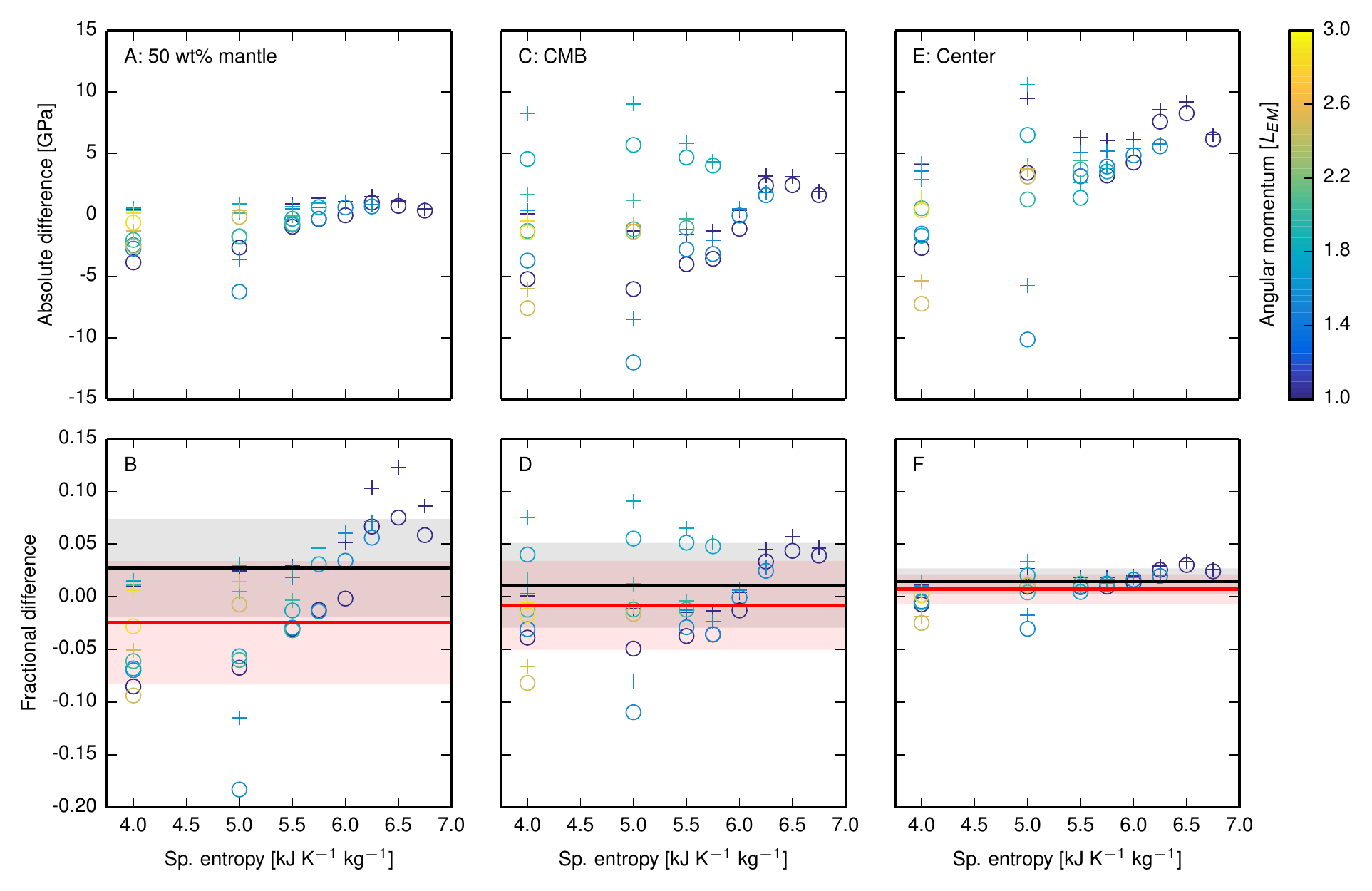}
\caption{Planetary structures calculated using SPH and HERCULES have similar internal pressures. Rows show the absolute (top) and fractional (bottom) difference in pressures between SPH and HERCULES calculations. Columns show the pressure difference in different locations in the Earth-mass bodies with different angular momenta (colors) and isentropic mantles of varying specific entropy (x-axis). Comparisons were made to HERCULES planets with a bounding pressure of both 10~bar ($+$) and a pressure equivalent to the lowest pressure in the midplane of the SPH structure ($\circ$). The solid black and red lines in the bottom row show the mean fractional pressure differences using HERCULES planets with 10~bar and the maximum SPH pressure respectively. The shaded area shows one standard deviation in the fractional errors. The shape and pressure contours for the SPH bodies plotted here are shown in Figure~4 of \cite{Lock2017}.
}
\label{sup:fig:HERCULES_comparison}
\end{sidewaysfigure*}

\clearpage

\begin{table}
\caption[SPH simulation results]{Summary of SPH impact simulations used in this paper and properties of their post-impact states. Most calculations were reported in \cite{Lock2017} but additional simulations of near mutual escape velocity collisions with high impactor to target mass ratios have also been performed. For each impact, the table includes: an index number; target mass $M_{\rm t}$; number of SPH particles in target, $N_{\rm t}$; target equatorial radius, $R_{\rm t}$; target angular momentum, $L_{\rm t}$; target 50\% mantle pressure, $p_{50}^{\rm t}$;  target core-mantle boundary (CMB) pressure, $p_{\rm CMB}^{\rm t}$;  target central pressure, $p_{\rm center}^{\rm t}$; projectile mass, $M_{\rm p}$; number of SPH particles in projectile, $N_{\rm p}$; projectile radius, $R_{\rm p}$; projectile angular momentum, $L_{\rm p}$; impact velocity, $V_{\rm i}$; impact parameter, $b$; modified specific energy, $Q_{\rm S}$; final simulation time; bound mass of post-impact structure, $M_{\rm bnd}$; core mass fraction of post-impact body, $f_{\rm core}$; bound mass angular momentum, $L_{\rm bnd}$; angular velocity of the dense ($\rho$~$>$~$1000$~kg~m$^{-3}$) region of post-impact structure, $\omega_{\rho}$; spin period of dense region, $T_{\rho}$; moment of inertia of bound mass, $I_{\rm bnd}$; post-impact 50\% mantle pressure, $p_{50}^{\rm bnd}$;  post-impact CMB pressure, $p_{\rm CMB}^{\rm bnd}$;  post-impact central pressure, $p_{\rm center}^{\rm bnd}$; 50\% mantle pressure in a body of the same mass, core fraction, and angular momenta but with a hot partially-vaporized, thermally-stratified thermal structure with an upper mantle entropy such that the body is at the CoRoL, $p_{50}^{\rm CoRoL}$;  CMB pressure at the CoRoL, $p_{\rm CMB}^{\rm CoRoL}$;  central pressure at the CoRoL, $p_{\rm center}^{\rm CoRoL}$; angular velocity of a magma-ocean planet with the same mass, angular momentum and core-mass fraction as the post-impact body, $\omega_{\rm MO}$; moment of inertia of the corresponding magma-ocean planet, $I_{\rm MO}$; 50\% mantle pressure in corresponding magma-ocean planet, $p_{50}^{\rm MO}$;  CMB pressure in corresponding magma-ocean planet, $p_{\rm CMB}^{\rm MO}$;  central pressure in corresponding magma-ocean planet, $p_{\rm center}^{\rm MO}$; 50\% mantle pressure of a magma-ocean planet with the same mass and core-mass fraction as the post-impact body but with an angular momentum corresponding to that of the Earth's when the lunar orbit reached the Cassini-state transition, $p_{50}^{\rm Cassini}$;  CMB pressure in corresponding magma-ocean planet at the Cassini state transition, $p_{\rm CMB}^{\rm Cassini}$;  central pressure in corresponding magma-ocean planet at the Cassini-state transition, $p_{\rm center}^{\rm Cassini}$; and post-impact structure dynamical class.
}
\label{sup:tab:impacts}
\end{table}


\end{document}